\documentclass[preprint,aps,eqsecnum]{revtex4}
\newcommand{\pslush}{p\!\!\!/}
\newcommand{\be}{\begin{equation}}
\newcommand{\bea}{\begin{eqnarray}}
\newcommand{\ee}{\end{equation}}
\newcommand{\eea}{\end{eqnarray}}

\usepackage{axodraw}
\usepackage{amssymb}
\usepackage{amsfonts}
\usepackage{pstricks}
\usepackage{graphicx}

\def\chic#1{{\scriptscriptstyle #1}}

\begin{document}

\title{The neutrino charge radius is a physical observable}

\author{J. Bernab\'eu, J. Papavassiliou, and J. Vidal}

\address{
Departamento de F\'{\i}sica Te\'orica and IFIC, \\
Universidad de Valencia-CSIC, \\
E-46100, Burjassot, Valencia, Spain }

\preprint{FTUV-02-1003, IFIC/02-47}

\thispagestyle{empty}

\begin{abstract}

We present  a method which allows,  at least in  principle, the direct
extraction  of the  gauge-invariant  and process-independent  neutrino
charge  radius  (NCR)   from  experiments.   Under  special  kinematic
conditions,  the judicious combination  of neutrino  and anti-neutrino
forward  differential  cross-sections  allows  the  exclusion  of  all
target-dependent contributions,  such as gauge-independent box-graphs,
not  related to  the NCR.   We  show that  the remaining  contributions
contain universal, renormalization  group invariant combinations, such
as  the electroweak  effective charge  and the  running  mixing angle,
which  must be  also separated  out.  By  considering  the appropriate
number of independent experiments  we show that one may systematically
eliminate these universal terms,  and finally express the NCR entirely
in  terms  of  physical  cross-sections.  Even  though  the  kinematic
conditions  and  the  required   precision  may  render  the  proposed
experiments unfeasible, at the conceptual level the analysis presented
here  allows for  the promotion  of the  NCR into  a  genuine physical
observable.

\medskip


\end{abstract}

\maketitle

\newpage

\section{Introduction}\label{sec:Int}

The diagrammatic definition  of off-shell electromagnetic form-factors
in the context of non-Abelian  gauge theories is known to be different
from the scalar or QED  cases, in the sense that the ``single-photon''
approximation gives rise  to gauge-dependent, and therefore unphysical
results  \cite{Fujikawa:1972fe,Abers:1973qs}.   
The  neutrino  electromagnetic
form-factor  has  been  a  celebrated  example of  this  general  fact
\cite{Papavassiliou:ex}.   Within  the  Standard Model  the  effective
photon-neutrino  interaction   generated  through  one-loop  radiative
corrections is  expected to  give rise to  a non-zero  neutrino charge
radius (NCR) \cite{Bernstein:jp,Bardeen:1972vi,Lee:1972fw,Lee:1977ti,
Dolgov:1981hv}, which, heuristically speaking, has
been traditionally associated with the electromagnetic ``size'' of the
neutrino \cite{SL}.  
The extraction of  this quantity from an off-shell one-loop
photon-neutrino vertex  $\Gamma^{\mu}_{A \nu_i \bar{\nu_i}}$  has been
carried out  in various gauge-fixing  schemes, leading to  the general
conclusion  that, in  the  absence of  a  definite guiding  principle,
important physical requirements  such as gauge-invariance, finiteness,
and    target-independence   
\cite{Lucio:1984mg,Monyonko:1984gb,Grau:1986cn,Auriemma:ak,Vogel:1989iv,
Degrassi:1989ip,Musolf:1991sa,Cabral-Rosetti:2000ad},    
could    not   be
simultaneously satisfied.  The non-trivial  task in this context is to
identify the correct  subset of Feynman graphs, which  would give rise
to a gauge-invariant and finite result for the NCR, while, at the same
time,  retaining  the  process-independence  of  the  definition.   In
particular, the most obvious manifestly gauge-invariant alternative of
computing the  entire process, and  then forcing the answer  to assume
the  form  of  a  ``single-photon'' interaction  
is  (by  definition)
process-dependent, because in the  computation of the entire amplitude
enter non-``single-photon'' contributions.  Therefore, adopting such a
procedure   precludes   the   conceptually  appealing   possibility   of
interpreting the resulting form-factor as an intrinsic property of the
particle in question. 

A definite  solution to  this problem has  recently been  presented in
\cite{Bernabeu:2000hf}. In  that work the  necessary guiding principle
is    provided    by    the    pinch    technique    (PT)    formalism
\cite{Cornwall:1982zr,Cornwall:1989gv,
Papavassiliou:1990zd,Degrassi:1992ue}, which implements the separation
of  a physical  amplitude into  electroweak  gauge-invariant effective
self-energy,   vertex   and   box  sub-amplitudes.    The   conceptual
requirement that  the effective  electromagnetic vertex of  a particle
has to be  process-independent, i. e., independent of  the target used
to probe the properties  of the particle, is automatically implemented
in this  construction.  In  particular, the NCR  is extracted  from an
effective  one-loop  proper  vertex  $\widehat{\Gamma}^{\mu}_{A  \nu_i
\bar{\nu_i}}$, which is independent of the gauge-fixing parameter, and
satisfies a QED-like Ward identity.  As has been demonstrated by means
of   detailed   calculations   in   \cite{Bernabeu:2000hf},   the   PT
construction   of    the   vertex   $\widehat{\Gamma}^{\mu}_{A   \nu_i
\bar{\nu_i}}$ amounts  to computing directly  the corresponding proper
vertex  in   the  Feynman  gauge   of  the  Background   Field  Method
\cite{Abbott:1980hw},   using    the   Feynman   rules    derived   in
\cite{Denner:1994xt}; this  fact is  in accordance with  the generally
known correspondence  between the PT and the  Background Field Method,
at  one \cite{Denner:1994xt,  Hashimoto:1994ct,  Pilaftsis:1996fh} and
two                                                               loops
\cite{Papavassiliou:2000az,Papavassiliou:2000bb,Binosi:2002bs}.

The  next important question  in this  context is  whether the  NCR so
defined constitutes  a genuine physical observable,  and in particular
how  it can  be extracted,  even in  principle, from  experiment.
The
general strategy  of how to  accomplish this from $\nu-e$ and 
$\nu-\nu$ cross-sections 
has  been  addressed  in a  recent  brief
communication \cite{Bernabeu:2002nw}.   
In this paper we present a detailed proof of the observable 
character of the NCR by its explicit separation from other 
renormalization group invariant (RGI) combinations in the physical 
cross-sections. 

As has been explained in  \cite{Bernabeu:2002nw}, 
the main difficulty one needs to
overcome in this context is the following: The PT rearrangement of the
$S$-matrix makes  possible the definition  of distinct sub-amplitudes,
which are individually  endowed with desirable theoretical properties;
one   of   these   sub-amplitudes,  $\widehat{\Gamma}^{\mu}_{A   \nu_i
\bar{\nu_i}}$, is directly related  to the NCR. However, the remaining
sub-amplitudes, even  though they do  no enter into the  definition of
the NCR, still contribute numerically to the entire $S$-matrix.  Thus,
in order  to extract the NCR,  one must conceive of  an experiment, or
combination of experiments, such that all contributions not related to
the NCR will be eliminated.

In this paper  we study in detail a set  of such experiments involving
neutrinos  and anti-neutrinos.   In  particular, we  elaborate on  the
``neutrino--anti-neutrino method'',  which allows for  the elimination
of the  box contributions.  The  general idea is to  study appropriate
combinations  involving  the   one-loop  cross-sections  of  (elastic)
processes of  the type  $ e \nu  \to e  \nu $ and  $e \bar{\nu}  \to e
\bar{\nu}$,  and  exploit  the  fact  that  the  box  diagrams  behave
differently  than vertex  or self-energy  diagrams under  the exchange
$\nu  \leftrightarrow   \bar{\nu}$,  or  equivalently,   under  charge
conjugation \cite{Sarantakos:1983bp}.   It turns  out that the  sum of
the total cross-sections of the  two processes mentioned above is free
of box  contributions.  This, together  with the fact that  the vertex
corrections  not related  to  the  NCR,
together with  the Bremsstrahlung  contributions vanish in  the special
kinematic limit of  zero momentum transfer, where the  NCR is actually
defined, allows for the isolation of three distinct parts: the desired
NCR, which  depends explicitly on the  flavour of the  neutrino one is
considering,  together with  two universal  parts,  i.e. contributions
that are completely flavour-  and target-independent, one consisting of
the tree-level and one-loop $Z$-boson propagator, and the other of the
one-loop mixing between $A$ and $Z$.  There is an
important theoretical difference however between the flavour-dependent
NCR  and the  two universal  pieces:  The NCR  is ultraviolet  finite,
whereas the  universal parts are ultraviolet divergent,  and they must
therefore be renormalized.  This  fact raises an important issue which
we will address in detail in this paper.

Specifically, in order to assign an observable character to individual
sub-amplitudes,  one needs  in  addition an  explicit separation  into
RGI quantities.   Otherwise  the
separation will  depend on the  way the renormalization  subtraction is
carried out,  i.e.  it  would be scheme-dependent.   In this  paper we
will employ the  concepts of the electroweak effective  charge, and of
the effective (running) electroweak mixing angle, in order to cast the
aforementioned    universal   contributions   into    manifestly   RGI
combinations
\cite{Hagiwara:1994pw,Papavassiliou:1996fn,Papavassiliou:1997pb}.    An
important consequence of  this analysis is in fact related
to the  very definition  of the NCR.   Specifically, the  universal PT
one-loop   $AZ$   self-energy   is   gauge-independent   and   couples
electromagnetically to  the target  fermions.  Therefore, it  could be
considered  as a flavour-independent  contribution to  the NCR,  to be
added to the flavor-dependent one  stemming from the proper vertex; in
fact this  point of view has  been often advocated  in the literature.
However,  the $AZ$  self-energy  is a  scheme-dependent and  therefore
unphysical contribution,  which, as such  cannot form part of  the NCR
definition.   Instead,  it must  be  appropriately  combined with  the
tree-level contribution mediated by  the $Z$-boson (which is certainly
not part  of the  NCR), in  order to form  the universal  RGI quantity
known as  the effective (running) electroweak mixing  angle.  Once the
decomposition of the result into RGI quantities has been accomplished,
one may  proceed unambiguously into their  experimental separation, by
considering the appropriate number  of different processes.  Thus, the
Standard Model prediction for the NCR, together with the two universal
RGIs,  can be  finally  expressed individually  in  terms of  specific
combinations of physical cross-sections.

The paper  is organized as follows: In  section~\ref{sec:PT} we review
the PT rearrangement of  the elastic scattering amplitude, focusing on
the relevant kinematic limit of vanishing momentum transfer, where the
NCR is defined. To simplify  the analysis we choose the charged target
fermions to  be right-handedly polarized  electrons, and we  show that
both the  vertex corrections  not related  to the NCR  as well  as the
Bremsstrahlung  corrections  vanish  in the  aforementioned  kinematic
limit.       In     section~\ref{sec:N-N}      we      present     the
neutrino--anti-neutrino  method in  detail, using  as  target fermions
both  right-handedly polarized  as well  as unpolarized  fermions.  In
section~\ref{sec:RGI} we review  the renormalization properties of the
relevant one-loop PT self-energies  appearing in the neutral sector of
the Standard  Model, and we show  how the resulting  amplitudes may be
written in terms of the  ultraviolet finite NCR and two manifestly RGI
building  blocks.   In section~\ref{sec:EXT}  we  present two  methods
which allow for  the individual extraction from an  appropriate set of
experiments of  the NCR and the  two RGI quantities  introduced in the
previous section.   In addition,  we present the  theoretical Standard
Model   predictions   for  these   three   quantities.   Finally,   in
section~\ref{sec:Con} we present our conclusions.

\setcounter{equation}{0}
\section{The  PT reorganized forward amplitude}\label{sec:PT}

In this section we will first review briefly some of the main results
presented in \cite{Bernabeu:2000hf} in an attempt to fix the notation 
and stress the relevant conceptual points. Then, we will show that 
in the special kinematic limit of zero momentum transfer, 
in which the NCR is in fact defined,  
the vertex corrections not related to the NCR, together with the 
Bremsstrahlung contributions, vanish.

For concreteness we will 
focus on  the process 
 $ e(k_1) \nu_{\mu} (p_1) 
\to e(k_2) \nu_{\mu} (p_2) $,
shown in Fig.1
The above process is chosen to be elastic with
the Mandelstam variables defined as
$s=(k_1+p_1)^2 = (k_2+p_2)^2$, 
$t= q^2 = (p_1-p_2)^2 = (k_1-k_2)^2$, 
$u = (k_1-p_2)^2 = (k_2-p_1)^2$, 
and $s+t+u=0$. 
The reason for considering  
$\nu_{\mu}$ instead of $\nu_{e}$ is because 
in this way one eliminates the charged channel mediated by a $W$-boson 
(Fig.1j).

The two relevant tree-level photon ($A$) and $Z$-boson 
vertices $\Gamma_{A {f} \bar{f}}^{\mu}$ and
$\Gamma_{{\chic Z} {f} \bar{f}}^{\mu}$ are given by
\bea
\Gamma_{A {f} \bar{f}}^{\mu} &=& -i e  
Q_{f} \gamma^{\mu} = -i e Q_{f} \gamma^{\mu} (P_{\chic L} + P_{\chic R})
\nonumber\\ 
\Gamma_{Z {f} \bar{f}}^{\mu} &=& \, 
-i \bigg(\frac{g_w}{c_w}\bigg)\, \gamma^{\mu}\, 
[ (s^2_w Q_{f} - T^{f}_z) P_{\chic L} + s^2_w Q_{f} 
P_{\chic R}]\nonumber\\
&=& \, 
-i \bigg(\frac{g_w}{c_w}\bigg)\, \gamma^{\mu}\, 
[ a_f + b_f \gamma_5]
\label{GenVer}
\eea
with 
$a_f = s^2_w Q_{f} - \frac{1}{2} T^f_z$  and 
$b_f=\frac{1}{2} T^f_z$.
In the above formulas 
$Q_f$ is the electric charge of the fermion $f$, 
$T^f_z$ its $z$-component of the weak iso-spin,  and
$P_{\chic R(\chic L)} = [1  + (-) \gamma_5]/2$  is the  chirality projection
operator, 
$c_w = \sqrt{1 - s^2_w} = M_{\chic W}/M_{\chic Z}$, and 
the electric charge $e$ is related to the $SU(2)_L$ 
gauge coupling $g_w$ by $e=g_w s_w$.

\begin{figure}[!t]
\includegraphics[width=15.0cm]{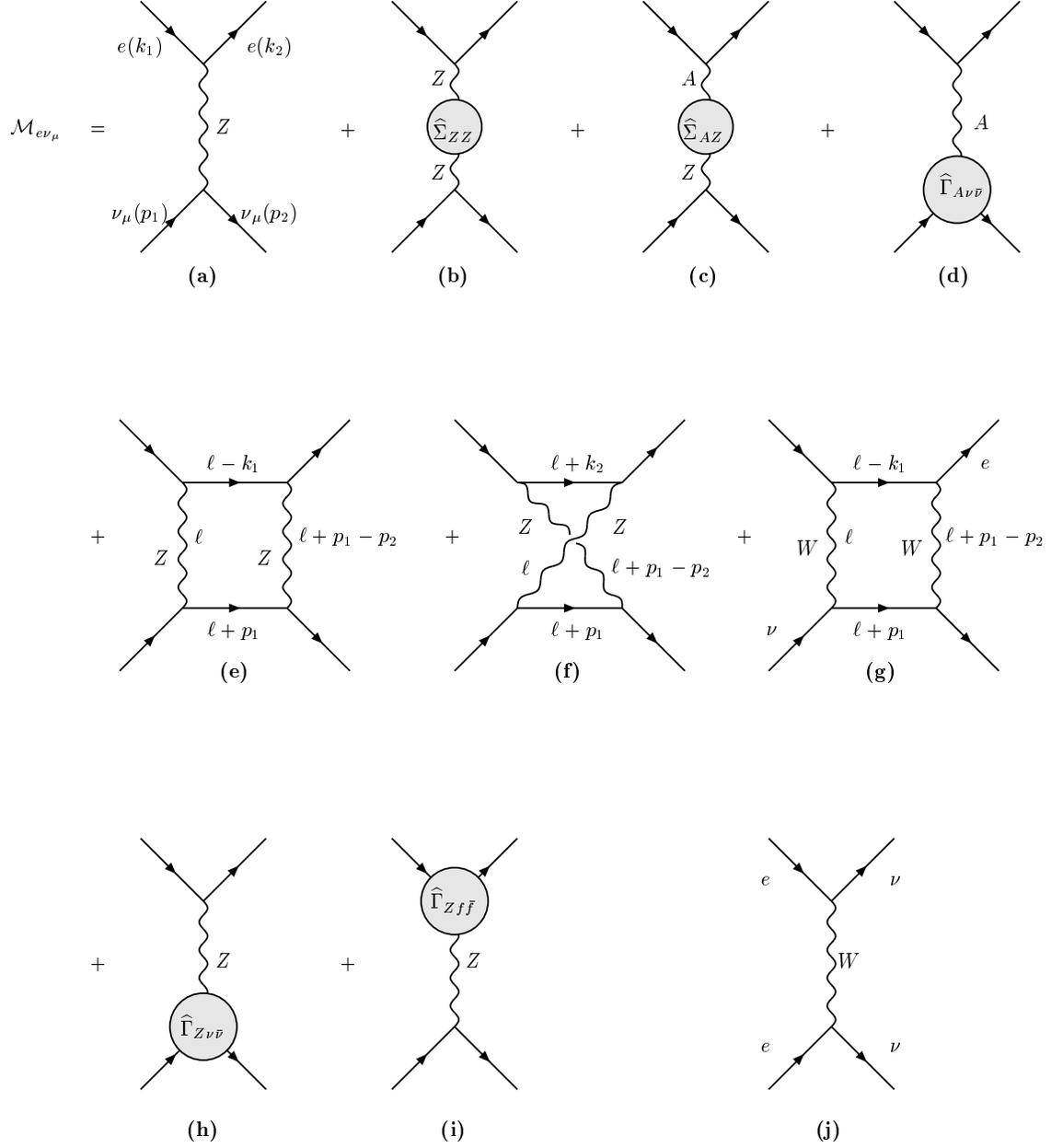}
\caption{\label{Fig1}   
The various classes of diagrams contributing to the one-loop amplitude
($\mathbf{a}$--$\mathbf{i}$). The charged channel ($\mathbf{j}$) 
vanishes for neutrino flavours other than $\nu_e$.}
\end{figure}

The one-loop contributions to the amplitude are shown
in Fig.1b -- Fig.1i . 
We will assume throughout that the  PT rearrangement of the
amplitudes has been carried out, exactly as described 
in  \cite{Bernabeu:2000hf}. In particular, 
after a series of crucial gauge-cancellations enforced by
the elementary Ward identities of the theory
\cite{Cornwall:1982zr,Cornwall:1989gv,Papavassiliou:1990zd}, 
the 
amplitude has been split into individually gauge-invariant
sub-amplitudes which correspond kinematically to
self-energies, vertices, and boxes. As has been explained in detail
in the literature \cite{Denner:1994xt, Hashimoto:1994ct, Pilaftsis:1996fh}, 
these latter  PT quantities coincide with the 
corresponding Green's functions computed in the 
framework of the Background Field Method, at the special 
value of the (quantum) gauge fixing parameter $\xi_Q=1$. 
Notice that the gauge-invariant  
``pure'' box contributions  
coincide with 
the conventional box contributions computed 
in the renormalizable Feynman gauge 
($R_{\xi}$ gauges, with $\xi =1$).

It is well-known \cite{Watson:1994tn}
that the  PT rearrangement of the amplitude may be carried out regardless of 
the kinematical details, as for example the specific values of
the Mandelstam variables, or the masses of the external particles.  
In what follows we will consider the above amplitude in  
the zero transfer limit, 
$t=q^2 \to 0$, $s=-u$, where the NCR is actually defined. 
In addition, we 
will assume that all external (on-shell) fermions 
are massless. 
As a result of this special kinematic situation we have the
following relations:
\bea
p_1^2 &=& p_2^2 = k_1^2 = k_2^2 = p_1 \cdot p_2 = k_1 \cdot k_2 = 0 
\nonumber\\
p_1 \cdot k_1  &=& p_1 \cdot k_2 = p_2 \cdot k_1 = p_2 \cdot k_2 = s/2 . 
\label{KINREL}
\eea
As we will see in a moment, in the aforementioned limit 
of $q^2 \to 0$, 
the one-loop  PT   
vertex corrections 
to the $Zf\bar{f}$ and $Z \nu \bar\nu{}$
vertices vanish, and so do the Bremsstrahlung 
contributions. Moreover, the special kinematic relations given in 
Eq.(\ref{KINREL}) are crucial for the validity of the 
neutrino--anti-neutrino method which we will present in the 
next section. 

In the center-of-mass system we have that 
$t=-2 E_{\nu}E_{\nu}'(1-x)\leq 0 $, 
where $E_{\nu}$ and $E_{\nu}'$
are the energies of the neutrino before and after the
scattering, respectively, and 
$x \equiv \cos\theta_{cm}$, where
$\theta_{cm}$
is the scattering angle. Clearly, the condition $t=0$ 
corresponds to the exactly forward amplitude, 
with $\theta_{cm}=0$, \, $x=1$. 
Equivalently, in the laboratory frame,
where the (massive) target fermions are at rest, the 
condition of $t=0$ corresponds to the kinematically extreme 
case where the target 
fermion remains at rest after the scattering.  

\begin{figure}[!t]
\includegraphics[width=15.0cm]{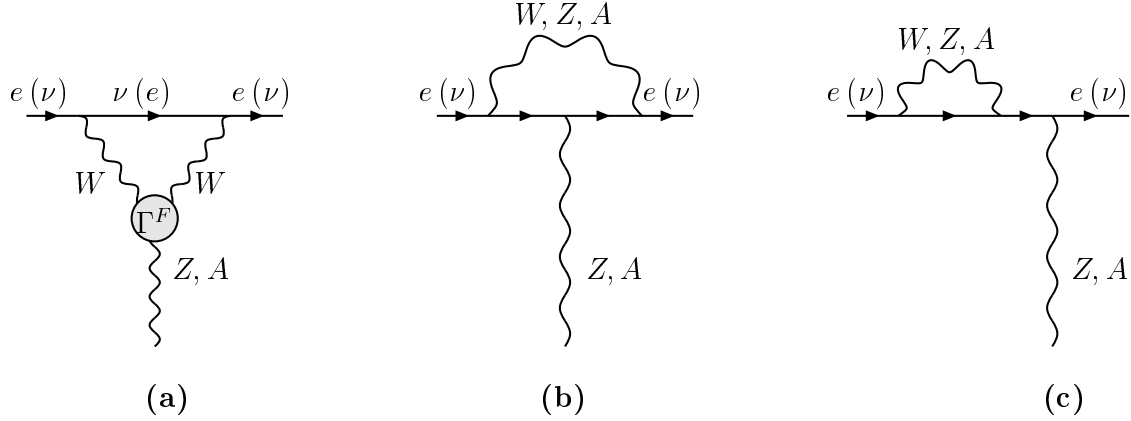}
\caption{\label{Fig2}   
The relevant vertex graphs, ($\mathbf{a}$)--($\mathbf{b}$),
and the fermion wave-function correction, ($\mathbf{c}$).}
\end{figure}

The relevant quantities 
which will appear in our calculations are the $ZZ$ and $AZ$
self-energies, to be denoted by 
$\widehat{\Sigma}_{\chic{Z}\chic{Z}}^{\mu\nu}(q^2)$ and 
$\widehat{\Sigma}_{\chic{A}\chic{Z}}^{\mu\nu}(q^2)$, respectively,
and three one-loop vertices 
$A \nu_i \bar{\nu_i} $, 
$Z\nu_i \bar{\nu_i}$, and $Z f \bar{f}$, to be denoted by 
$\widehat{\Gamma}^{\mu}_{A \nu_i \bar{\nu}_i}$, 
$\widehat{\Gamma}^{\mu}_{Z \nu_i \bar{\nu}_i}$, and 
$\widehat{\Gamma}^{\mu}_{Z f \bar{f}}$, respectively.
In the   PT framework 
$\widehat{\Sigma}_{\chic{A}\chic{Z}}^{\mu\nu}(q^2)$
is transverse, for {\it both} 
the fermionic and the bosonic contributions,
i.e. it may be written in terms of the 
dimension-less scalar function 
${\widehat{\Pi}}_{ \chic{A} {\chic Z}} (q^2)$ as
\be
\widehat{\Sigma}_{\chic{A}\chic{Z}}^{\mu\nu}(q^2)
= 
\Bigg(  q^2 \, g^{\mu\nu}  - q^{\mu} q^{\nu}\Bigg) 
{\widehat{\Pi}}_{ \chic{A} {\chic Z}} (q^2)\, .  
\label{d4}
\ee
On the other hand, $\widehat{\Sigma}_{\chic{Z}\chic{Z}}^{\mu\nu}(q^2)$
is of course not transverse. In what follows we will 
discard all longitudinal pieces, since they vanish between 
the conserved currents of the massless external fermions,
and will 
keep only the part proportional to $g^{\mu\nu}$; 
its dimension-full cofactor will be denoted 
by $\widehat{\Sigma}_{\chic{Z}\chic{Z}}(q^2)$, i.e.
\be
\widehat{\Sigma}_{\chic{Z}\chic{Z}}^{\mu\nu}(q^2)
= \widehat{\Sigma}_{\chic{Z}\chic{Z}}(q^2) g^{\mu\nu}
\ee
The closed one-loop expressions for 
$\widehat{\Sigma}_{\chic{Z}\chic{Z}}^{\mu\nu}(q^2)$ and 
$\widehat{\Sigma}_{\chic{A}\chic{Z}}^{\mu\nu}(q^2)$
can be found in various places in the literature 
\cite{Degrassi:1992ue,Degrassi:1993kn,Hagiwara:1994pw}.

It is relatively easy to convince oneself 
that, if one were to relax the masslessness condition 
 for external (target) fermions,
 all additional 
contributions due to their non-vanishing 
masses $m$   
always appear proportional to positive powers of $(m/M)$ and/or
$(m/\sqrt{s})$, where 
$M$ stands for the mass 
of the $W$ or $Z$ bosons. 
Clearly, the terms  $(m/M)$ are 
naturally suppressed because of the heaviness of the gauge bosons.
On the other hand, the terms $(m/\sqrt{s})$ can be made arbitrarily small,
by letting the variable 
$s$, which in principle 
can be controlled by adjusting the energies of the 
incoming particles,  
reach sufficiently high values.

Turning to the vertex corrections, 
the NCR, to be denoted by $\big <r^2_{\nu_i}\,  \big>$,
will be defined from the vertex 
$\widehat{\Gamma}^{\mu}_{A \nu_i \bar{\nu}_i}$, which 
is given by the two 
graphs of Fig.2a and  Fig.2b (with a 
photon $A$ and neutrinos $\nu$ entering into the vertex).
As has been explained in detail in the literature, 
the construction of the
gauge independent and gauge invariant one-loop vertex by means of the
 PT finally amounts to 
using the Feynman gauge for the all gauge-boson propagators,
and replacing 
the usual three-boson vertex
\be
\Gamma_{\alpha\mu\nu} (q,k,-k-q) 
= (q-k)_{\nu} g_{\alpha\mu} + (2k+q)_{\alpha} g_{\mu\nu} 
- (2q+k)_{\mu} g_{\alpha\nu} 
\ee
by the tree-level vertex 
\be
\Gamma_{\alpha\mu\nu}^{\chic F} = 
(2k+q)_{\alpha} g_{\mu\nu} + 2q_{\nu}g_{\alpha\mu} 
- 2q_{\mu}g_{\alpha\nu}\, .
\label{GF}
\ee
The vertex $\Gamma_{\alpha\mu\nu}^{\chic F}$ 
satisfies the elementary Ward identity
\be 
q^{\alpha} \Gamma_{\alpha\mu\nu}^{\chic F}= 
 (k+q)^2 g_{\mu\nu} - k^2  g_{\mu\nu}\, ,
\label{WI2B}
\ee
Equivalently, one may use directly the 
Feynman rules of the Background Field Method \cite{Denner:1994xt}, 
choosing for the gauge-fixing parameter $\xi_Q$
of the (quantum) bosons the value $\xi_Q = 1$.

It is straightforward to evaluate
the two aforementioned vertex graphs; their  
sum gives a ultra-violet finite result, from which 
one can extract the 
dimension-less
electromagnetic form-factor $\widehat{F}_1^{A}(q^2)$.  
In particular, since $\widehat{F}_1^{A}(q^2)$ is 
proportional to $q^2$, we may define the dimension-full
form-factor $\widehat{F}_{\nu_i}(q^2)$ as
\be
\widehat{\Gamma}^{\mu}_{A \nu_i \bar{\nu}_i}
= \widehat{F}_1^{A}(q^2) \bigg[ie \gamma_{\mu}(1-\gamma_{5})\bigg]\,
= q^2  \widehat{F}_{\nu_i}(q^2)
\bigg[ie \gamma_{\mu}(1-\gamma_{5})\bigg]
\label{TheFS}
\ee
$\widehat{F}_{\nu_i}(q^2)$ depends on the mass $m_i$ of the
charged iso-doublet
partner of the neutrino, which appears in the two relevant 
Feynman diagrams.
In the limit of both
$q^2, m_i^2 \to 0$, $\widehat{F}_{\nu_i}(q^2)$  is infrared divergent,
whereas it is infrared finite in the limit
 $q^2\to 0$, $m_i^2 \neq 0$. 
After canceling the $q^2$ 
against the 
photon propagator, we can take the
limit $q^2 \to 0$, keeping  $m_i^2$ non-zero.
Defining as usual 
$ \widehat{F}_{\nu_i}(0) = \frac{1}{6} \big <r^2_{\nu_i}\,  \big>$, 
we finally arrive at \cite{Bernabeu:2000hf}
\be
\big <r^2_{\nu_i}\,  \big> =\, 
\frac{G_{\chic F}}{4\, {\sqrt 2 }\, \pi^2} 
\Bigg[3 
- 2\log \Bigg(\frac{m_{\chic i}^2}{M_{\chic W}^2} \Bigg) \Bigg]\, ,
\,\,\,\,\,\,\,\, i= e,\mu,\tau
\label{ncr}
\ee
where $G_{\chic F} = g_w^2 \sqrt{2}/8 M_{\chic W}^2 $ 
is the Fermi constant.
Notice that the logarithmic term in the above expression
originates entirely from the Abelian-like diagram of Fig.2b.

\subsection{Vanishing of the $Zee$ and $Z\nu\nu$ vertex corrections}

We will show that the sum of the 
one-loop vertex and wave-function
corrections, which are collectively depicted in
Fig.1h and Fig.1i vanishes in the limit $q^2 \to 0$.
Since these diagrams are multiplied 
by a massive tree-level $Z$ propagator $D_{\chic Z}(q)$,
which is regular (non-divergent) in this limit,
they do not contribute to the scattering amplitude we consider.
This is to be contrasted with the 
$\widehat{\Gamma}^{\mu}_{A \nu_i \bar{\nu}_i}$,  
which is accompanied by a 
$(1/q^2)$ photon-propagator, thus giving rise 
to a contact interaction between the target-fermion and the neutrino,
described by the NCR. 

It is known \cite{Papavassiliou:1990zd}
that the one-loop PT vertex 
$\widehat\Gamma_{{\chic Z} {\chic f} {\chic f}}^{\alpha}(q,p_1,p_2)$
with $f = e$  or $f = \nu$, shown in Fig.2a and Fig.2b, 
satisfies a QED-like 
Ward identity, relating it to the 
PT inverse fermion propagators $\widehat\Sigma_f$,  shown in Fig.2c ,
i.e
\be 
q_{\alpha} 
\widehat\Gamma_{{\chic Z} {\chic f} {\chic f}}^{\alpha}(q,p_1,p_2)
= 
\widehat\Sigma_f (p_1) - \widehat\Sigma_f (p_2)
\label{QEDWI}
\ee 
Eq.(\ref{QEDWI})
is a straightforward consequence of the tree-level Ward identity 
of Eq.(\ref{WI2B}).
By virtue of Eq.(\ref{QEDWI}),
when the proper vertex graphs are combined
with the renormalization of the external fermions, the 
net result is ultraviolet finite 
(because, as in QED, $Z_1=Z_2$).
Below we give the results of the individual graphs
contributing to vertex
$\widehat\Gamma_{{\chic Z} {\chic f} \bar{\chic f}}^{\alpha}$, 
corresponding to the vertex graphs of Fig.2, 
with a $Z$-boson entering into the vertex.
These graphs are calculated in the limit where 
$t = q^2 \to 0 $, the fermions appearing in the
loop are considered strictly massless, and terms proportional
to $q^{\alpha}$ vanish, because 
they are contracted with a conserved current, since
the external fermions are considered massless as well. 
The graph in Fig.2b containing
the virtual photon is infrared divergent, even if the fermions 
are massive. We will regulate this divergence
by introducing a fictitious photon mass
$m_{\chic A}$, a procedure which is compatible with 
gauge-invariance. Equivalently one may use dimensional 
regularization to regularize both ultraviolet  and infrared 
divergences  \cite{Gastmans:uv,Marciano:tv}.
In any case, all contributions will cancel
algebraically, before the infrared cutoff is removed. 
The two quantities which naturally appear when calculating the
diagrams using standard techniques, such
as Feynman parametrization and dimensional integration, 
are the following:
\bea
I_1 (M^2) &=& \frac{1}{2} (-2 + \epsilon)^2 
\int_0^1 dx \, (1-x)\, 
\bigg[ \, \frac{2}{\epsilon} - \ln(x M^2)\bigg] \nonumber\\
I_2 (M^2) &=& (-2 + \epsilon)
\int_0^1 dx \,x \,\bigg[\, \frac{2}{\epsilon} - \ln(x M^2)\bigg] 
\eea
where $d=4-\epsilon$.
It is elementary to verify that $I_1 (M^2) + I_2 (M^2) = 0$;
notice that this relation holds not only for the divergent parts,
but also for the parts that 
are finite and non-vanishing as $\epsilon \to 0$.
In terms of $I_1$ and $I_2$ we have 
(we suppress a common factor $g_w^3 /16 \pi^2 c_w^3 $): 
\bea
\left[\bf{2b}\right]_{\chic A}^{(e)} &=& s_w^2 c_w^2
I_1 (m_{\chic A}^2) \,\gamma^{\alpha}(a_e+b_e\,\gamma_5)\,,  
\nonumber\\
\left[\bf{2c}\right]_{\chic A}^{(e)} &=& s_w^2 c_w^2 I_2 (m_{\chic A}^2) \,
\gamma^{\alpha}(a_e+b_e\,\gamma_5) \,, \\   
\left[\bf{2b}\right]_{\chic Z}^{(e)} &=& I_1 (M_{\chic Z}^2)\,
\gamma^{\alpha} 
\bigg[ a_e (a^2_e + 3 b^2_e) + b_e (b^2_e + 3 a^2_e)\gamma_5 \bigg]\,,
\nonumber\\
\left[\bf{2c}\right]_{\chic Z}^{(e)} &=& 
I_2 (M_{\chic Z}^2)\, \gamma^{\alpha}  
\bigg[a_e (a^2_e + 3 b^2_e) + b_e (b^2_e + 3 a^2_e)\gamma_5 \bigg]\,,\\
\left[\bf{2b}\right]_{\chic Z}^{(\nu)} &=& I_1 (M_{\chic Z}^2) \, 
\gamma^{\alpha} P_{\chic L} \,,\nonumber\\
\left[\bf{2c}\right]_{\chic Z}^{(\nu)} &=& I_2 (M_{\chic Z}^2) \,
\gamma^{\alpha} P_{\chic L} \,,\\
\left[\bf{2a}\right]^{(e)}_{\chic W \chic W} &=& - \frac{c_w^4}{2} \, 
I_2 (M_{\chic W}^2)\gamma^{\alpha} P_{\chic L} \,,\nonumber\\
\left[\bf{2b}\right]_{\chic W}^{(e)} &=&  
\frac{c_w^2}{4}\,I_1 (M_{\chic W}^2)\,  
\gamma^{\alpha} P_{\chic L}\,,\nonumber\\
\left[\bf{2c}\right]_{\chic W}^{(e)} &=& - \frac{c_w^2}{2}\, 
(a_e-b_e) \,I_2 (M_{\chic W}^2)\, 
\gamma^{\alpha} P_{\chic L} \,,\\
\left[\bf{2a}\right]^{(\nu)}_{\chic W \chic W} &=& \frac{c_w^4}{2} \, 
I_2 (M_{\chic W}^2)\,\gamma^{\alpha} P_{\chic L}\,, 
\nonumber\\
\left[\bf{2b}\right]_{\chic W}^{(\nu)} &=&
\frac{c_w^2}{2}\, (a_e-b_e)\, I_1 (M_{\chic W}^2)\,  
\gamma^{\alpha} P_{\chic L} \,,\nonumber\\
\left[\bf{2c}\right]_{\chic W}^{(\nu)} &=& -
\frac{c_w^2}{4}\, I_2 (M_{\chic W}^2) \,\gamma^{\alpha} P_{\chic L}\,.
\eea
where the subscripts on the left hand-side denote the virtual 
gauge boson(s) appearing inside the corresponding graphs,
and the superscripts specify the type of incoming fermion. 
It is straightforward to see that 
$\left[\bf{2b}\right]_{\chic A}^{(e)}+ \left[\bf{2c}\right]_{\chic A}^{(e)} =
\left[\bf{2b}\right]_{\chic Z}^{(e)}+ \left[\bf{2c}\right]_{\chic Z}^{(e)} = 
\left[\bf{2b}\right]_{\chic Z}^{(\nu)} + 
\left[\bf{2c}\right]_{\chic Z}^{(\nu)} =
\left[\bf{2a}\right]^{(e)}_{\chic W \chic W} + 
\left[\bf{2b}\right]_{\chic W}^{(e)}+
\left[\bf{2c}\right]_{\chic W}^{(e)}=
\left[\bf{2a}\right]^{(\nu)}_{\chic W \chic W}+
\left[\bf{2b}\right]_{\chic W}^{(\nu)}+ 
\left[\bf{2c}\right]_{\chic W}^{(\nu)} =0 $.  
To prove the cancellations we have also used 
that $a_e-b_e = -\frac{1}{2} +  c_w^2$.

\subsection{Vanishing of the Bremsstrahlung }

In this subsection we will show that 
the differential cross-section corresponding to the
Bremsstrahlung diagrams vanishes in the kinematic
limit of zero momentum transfer \cite{Passera:2000ug}.
The two diagrams contributing to the Bremsstrahlung process
$e(k_1)\, \nu_{e}(p_1) \to e(k_2) \,\nu_{e}(p_2) \,A (k_3)$
and 
$e(k_1)\, \nu_{e}(p_1)\, A (k_3) \to e(k_2) \,\nu_{e}(p_2)$
are shown in Fig.3. The
conservation of four-momentum assumes the form 
$k_1 + p_1 = k_2 + p_2 + k_3$, and 
$t = (p_1-p_2)^2 = (k_1-k_2-k_3)^2$.
A direct consequence of this special kinematic choice 
$t\to 0 $ are the relations 
\be
k_1 \cdot k_2 + k_1 \cdot k_3 -  k_2 \cdot k_3 = 0 \, ,   
\,\,\,\,\,
p_i \cdot ( k_2 +   k_3 - k_1) = 0 \, , \,\,\,\, i=1,2
\ee

\begin{figure}[!t]
\includegraphics[width=15.0cm]{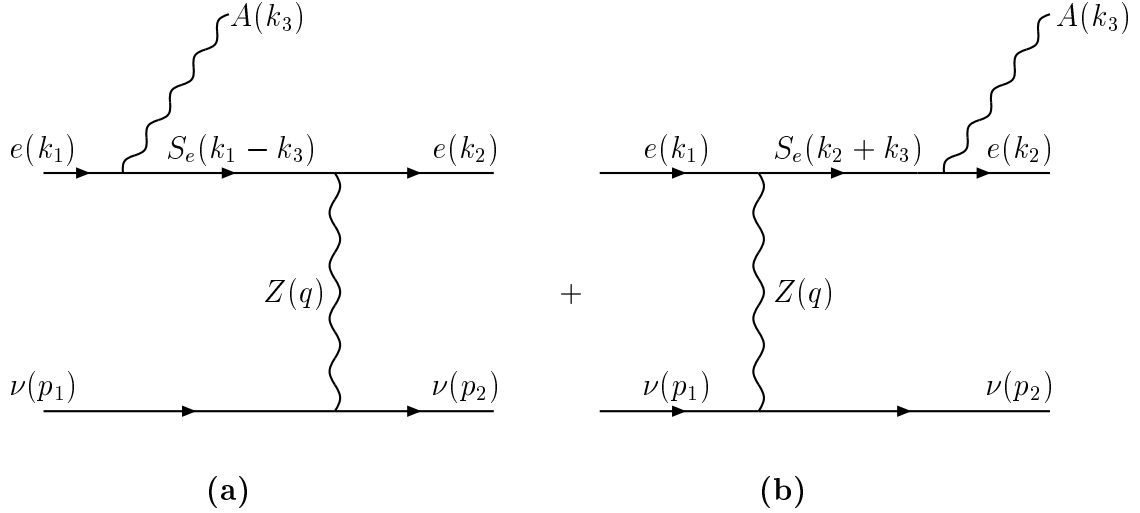}
\caption{\label{Fig3}   
The Bremsstrahlung diagrams}
\end{figure}

The $S$-matrix element 
${\cal M}^{\chic B}_{\mu}$ 
consists of the two parts,
${\cal M}_{a \,\mu}^{\chic B}$ and 
${\cal M}_{b \,\mu}^{\chic B}$, corresponding to the 
diagram (a) and diagram (b) of Fig.3, respectively, i.e.
\be
{\cal M}^{\chic B}_{\mu} 
= {\cal M}_{a \,\mu }^{\chic B} + {\cal M}_{b \,\mu }^{\chic B}
\ee
with
\bea
{\cal M}_{a \,\mu}^{\chic B} &=&  
[\bar{u}(k_2)\gamma_{\rho}(a_{e}+b_{e}\gamma_{5}) S_e (k_1-k_3)
\gamma_{\mu} u(k_1)]
[\bar{u}(p_2)\gamma^{\rho} P_{\chic L} u(p_1)]
\nonumber\\
{\cal M}_{b \,\mu}^{\chic B} &=&
[\bar{u}(k_2)\gamma_{\mu} S_e (k_2 + k_3)
\gamma_{\rho} (a_{e}+b_{e}\gamma_{5}) u(k_1)]
[\bar{u}(p_2)\gamma^{\rho} P_{\chic L} u(p_1)]
\eea
where we have suppressed a common multiplicative
factor originating from the coupling constants.
The Bremsstrahlung differential cross-section, 
 $\kappa^{\chic B}$,  
is proportional to the square of the amplitude:
\bea
\kappa^{\chic B} &=& {\cal M}^{\chic B} * {\cal M}^{\chic B\dagger}
\nonumber\\
 &=& ({\cal M}_{a}^{\chic B} + {\cal M}_{b}^{\chic B})*
({\cal M}_{a}^{\chic B} + {\cal M}_{b}^{\chic B})^{\dagger}
\nonumber\\
 &=& {\cal M}_{a}^{\chic B}*{\cal M}_{a}^{\chic B\dagger} +
{\cal M}_{b}^{\chic B}*{\cal M}_{b}^{\chic B\dagger}
+ 2   \Re e ( {\cal M}_{a}^{\chic B}*{\cal M}_{b}^{\chic B\dagger})
\nonumber\\
 &\equiv & 
\kappa^{\chic B}_{a a} + \kappa^{\chic B}_{b b}
+2 \kappa^{\chic B}_{a b}
\label{SBREM}
\eea
The sum over polarizations for the photon introduces the 
usual polarization tensor 
$P^{\mu\nu}(k_3) = g^{\mu\nu} - 
(n^{\mu} k_3^{\nu} + n^{\nu} k_3^{\mu})/n\cdot k_3$, with
$n^{\mu}$ an arbitrary four-vector. 
Of course, $U(1)$ gauge invariance 
furnishes the tree-level Ward identity
$k_3^{\mu} {\cal M}^{\chic B}_{\mu} = 0$, so that
effectively only the $g^{\mu\nu}$ piece of  
$P^{\mu\nu}(k_3)$ contributes. 

Now for the electron propagators inside the above diagrams we use that
\be
S_e (k_1-k_3) = -\, \frac{\not\! k_1 -\not\! k_3}{2 k_1\cdot k_3} \, 
,
\,\,\,\,\,\, 
S_e (k_2+k_3) =  \frac{\not\! k_2 + \not\! k_3}{2 k_2 \cdot k_3} \, 
\ee
and so
\bea
\kappa^{\chic B}_{a a} &=& 
\frac{{\cal T}_{a a}}{(2\, k_1\cdot k_3)^2}  
\nonumber\\
\kappa^{\chic B}_{b b} &=&  \frac{{\cal T}_{b b}}{(2\, k_2\cdot k_3)^2}
\nonumber\\
\kappa^{\chic B}_{a b} &=& - \, 
\frac{{\cal T}_{a b}}{(2\, k_1\cdot k_3)(2\, k_2 \cdot k_3)}
\eea
with 
\bea
{\cal T}_{a a} &=&
\mbox{Tr}\bigg(\gamma_{\mu}\, 
(\not\! k_1 -\not\! k_3)\, 
\gamma_{\rho}\, (a_{e}+b_{e}\gamma_{5})\,
\not\! k_2\,
\gamma_{\sigma}\,(a_{e}+b_{e}\gamma_{5})\,
(\not\! k_1 -\not\! k_3)\,
\gamma_{\mu}\not\! k_1 \bigg)
\mbox{Tr}\bigg(\gamma^{\rho} \, P_{\chic L}
\not\! p_2 \, \gamma^{\sigma}\, 
P_{\chic L} \, \not\! p_1 \bigg)
\nonumber\\
{\cal T}_{b b} &=&
\mbox{Tr}\bigg(
\gamma_{\rho} \,(a_{e}+b_{e}\gamma_{5})\,
(\not\! k_2 +\not\! k_3)\, 
\gamma_{\mu} \,  \not\! k_2 \, \gamma_{\mu}\, 
(\not\! k_2 +\not\! k_3)\, 
\gamma_{\sigma} \,(a_{e}+b_{e}\gamma_{5})\, 
\not\! k_1 \bigg)
\mbox{Tr}\bigg(\gamma^{\rho} \, P_{\chic L}
\not\! p_2 \, \gamma^{\sigma}\, 
P_{\chic L} \, \not\! p_1 \bigg)
\nonumber\\
{\cal T}_{a b} &=& 
\mbox{Tr}\bigg(
\gamma_{\rho}\, (a_{e}+b_{e}\gamma_{5})\,
(\not\! k_2 +\not\! k_3)\, 
\gamma_{\mu} \,  \not\! k_2 \,
\gamma_{\sigma}\, (a_{e}+b_{e}\gamma_{5})\,
(\not\! k_1 -\not\! k_3)\,
\gamma_{\mu} \, \not\! k_1 \bigg)
\mbox{Tr}\bigg(\gamma^{\rho} \, P_{\chic L}
\not\! p_2 \, \gamma^{\sigma}\, 
P_{\chic L} \, \not\! p_1 \bigg) \nonumber\\
&&{}
\eea
One can then show, employing the standard properties of the 
trace of $\gamma$ matrices 
together with the special kinematic relations given
above, that 
\bea
{\cal T}_{a a} &=&
- 32 \, (2 k_1\cdot k_3) X_1 
\nonumber\\
{\cal T}_{b b} &=&
- 32 \,  (2 k_2\cdot k_3) X_2
\nonumber\\
{\cal T}_{a b} &=& 
- 16 \, \bigg[(2 k_1\cdot k_3) X_2 + (2 k_2\cdot k_3) X_1 \bigg] 
\eea
\bea
X_1 &=& 
(a_{e}-b_{e})^2 (p_2 \cdot k_2) (p_1 \cdot k_3) 
+ (a_{e}+b_{e})^2 (p_2 \cdot k_3) (p_1 \cdot k_2)
\nonumber\\
X_2 &=&
(a_{e}+b_{e})^2 (p_2 \cdot k_1) (p_1 \cdot k_3) 
+ (a_{e}-b_{e})^2 (p_2 \cdot k_3) (p_1 \cdot k_1)
\eea
Thus,
\bea
\kappa^{\chic B}_{a a} &=& 
- \frac{16 X_1}{ k_1\cdot k_3}  
\nonumber\\
\kappa^{\chic B}_{b b} &=& - \frac{16 X_2}{ k_2\cdot k_3}
\nonumber\\
\kappa^{\chic B}_{a b} &=& \frac{8 X_1}{\, k_1\cdot k_3}
+ \frac{8 X_2}{k_2\cdot k_3}
\eea
From these relations and Eq.(\ref{SBREM}) follows immediately that 
$ \kappa^{\chic B} = 0$, as announced; 
evidently, in the kinematic limit considered, a 
completely destructive interference takes place, 
which forces the cross-section
to vanish.
Notice that in arriving at the above result 
nowhere have we actually assumed that
the emitted photon is soft; in fact, the vanishing
of the cross-section has been shown simply by evaluating the 
traces, without need to enter into the specifics of the  
(implicit) three-body phase-space integration.

\section{The neutrino -- anti-neutrino method}\label{sec:N-N}

In this section we will present the neutrino -- anti-neutrino method
in detail. As we will see, this method isolates the 
process-dependent box contributions, at the expense of 
doubling the number of experiments needed. 

\subsection{Right-handed electrons}

We begin our analysis by choosing the target electrons to be 
right-handedly polarized. This choice 
eliminates the one-loop $WW$ box of Fig.1g, 
and makes the   
introduction to 
the neutrino--anti-neutrino method calculationally easier 
\cite{box}. 
In addition,  it will be used in section V as one of the processes 
that need be considered for the experimental extraction of the NCR.
We emphasize that 
this particular choice of target-fermions 
constitutes no loss of generality;
in fact, as we will see later in detail, all results 
derived using this particular process will   
be generalized after minor calculational adjustments, 
to the case of left-handed or unpolarized target fermions.

For the special case of 
right-handed fermions the two 
vertices of 
Eq.(\ref{GenVer}) assume the form
\bea
\Gamma_{A {f_R} \bar{f_R}}^{\mu} &=& -i e\,  
Q_{f} \gamma^{\mu} P_{\chic R} \nonumber\\ 
\Gamma_{Z {f_R} \bar{f_R}}^{\mu} &=& \, 
-i e\, Q_{f} \bigg(\frac{s_w}{c_w}\bigg) \,  
\gamma^{\mu}\, P_{\chic R}
\label{Verer}
\eea

The differential cross-section 
in the center-of-mass system is 
given by 
\be
\frac{d\sigma}{d\Omega_{cm}} = \frac{1}{64\pi^2 s} |{\cal M}|^2
\ee
where $d\Omega_{cm} = d x \, d \phi_{cm}$
and ${\cal M} $ is the amplitude.

For the tree-level amplitude ${\cal M}_{\nu_{\mu} e_{\chic R}}^{(0)}$ 
(Fig.1a), mediated only by an off-shell $Z$, we have
\be 
{\cal M}_{\nu_{\mu} e_{\chic{R}}}^{(0)} =
 c_{(a)}  A_{(a)} {\cal C}_{\chic L}
\ee
with
\be 
{\cal C}_{\chic L} = [\bar{u}_{\chic R}(k_2)\gamma_{\mu} u_{\chic R}(k_1)]
[\bar{u}(p_2)\gamma^{\mu} P_{\chic L} u(p_1)]\,,
\label{CL}
\ee
where $u_{\chic R} = P_{\chic R} u$, is the right-handed 
electron spinor, and
\be
 c_{(a)} = \frac{i e^2}{2 c^2_w} ~,~~~~~~~~~~
A_{(a)}   = - \frac{1}{M_Z^2} = \, D_{\chic Z}(0),
\ee 
where 
$D_{\chic Z}(q^2)= (q^2 -M_{\chic Z}^2)^{-1}$ is the scalar cofactor
of the tree-level $Z$ propagator 
$D_{\chic Z}^{\mu\nu}(q) = -i D_{\chic Z}(q^2) g^{\mu\nu}$.

At one-loop, 
\be
{\cal M}_{\nu_{\mu} e_{\chic{R}}}^{(1)} = 
\bigg(\sum_n c_n A_n\bigg) {\cal C}_{\chic L}
+ B_{\chic Z \chic Z} 
\ee
with $n=(b),(c),(d)$, namely the corresponding 
diagrams relevant for the process, shown in Fig.1 .    
The coefficients are given by 
\be
          c_{(b)} = \frac{i e^2}{2 c^2_w} ~, \,\,\,\,  
          c_{(c)} = \frac{i e g_w}{2 c_w} ~, \,\,\,\,
          c_{(d)} = 2 i e^2 ,            
\ee
and 
\bea
 A_{(b)}  &=& \frac{ -i \widehat{\Sigma}_{{\chic Z} 
                   {\chic Z}}(0)}{M_Z^4}\,,\nonumber\\
 A_{(c)}  &=&  \frac{ i \widehat{\Pi}_{{\chic A} 
                   {\chic Z}}(0)}{M_Z^2}\,, 
\nonumber\\
 A_{(d)} &=& \widehat{F}_{\nu_i}(0)\,.
\label{THEA}
\eea
Notice that the factor of 2 appearing in $c_{(d)}$ is due to the
fact that the usual definition of $\widehat{F}_1^{A}(q^2)$, 
given in Eq.(\ref{TheFS}),
involves $(1-\gamma_5)$ instead of the $P_{\chic L}$ which appears 
in ${\cal C}_{\chic L}$ of Eq.(\ref{CL})

Finally, $B_{\chic Z \chic Z}$ denotes  
the contribution of the 
box graphs shown in Fig.1e and Fig.1f,
\bea
B_{\chic Z \chic Z} 
&=& 
  - {\cal K}_{\chic Z\chic Z}  
[\bar{u}_{\chic R}(k_2)\gamma_{\rho} S_e (\ell-k_1)
\gamma_{\sigma}u_{\chic R}(k_1)]
[\bar{u}(p_2)\gamma^{\rho} P_{\chic L} S_{\nu_{\mu}} (\ell+p_1)
\gamma^{\sigma} P_{\chic L} u(p_1)]
\nonumber\\
&& + {\cal K}_{\chic Z\chic Z}  
[\bar{u}_{\chic R}(k_2)\gamma_{\rho} S_e (\ell+k_2)
\gamma_{\sigma}u_{\chic R}(k_1)]
[\bar{u}(p_2)\gamma^{\sigma} P_{\chic L} S_{\nu_{\mu}} (\ell+p_1)
\gamma^{\rho} P_{\chic L} u(p_1)] 
\label{BZZ}
\eea
with
\be
{\cal K}_{\chic Z\chic Z} 
\equiv \frac{e^4}{4\, c_w^4}
\int \frac{d^4\ell}{(2\pi)^4}
D_{\chic Z} (\ell)  D_{\chic Z} (\ell+p_1-p_2) 
\ee
where $\ell$
is the virtual four-momentum,
and $S(p) = i/\pslush$ is the 
massless tree-level fermion propagator. 
The relative minus sign in the contribution of the direct 
and crossed boxes originates from the fact that 
in the former graph the direction of 
the fermion propagator is opposite to the flow of the four-momentum
(see Fig.1e and  Fig.1f). 

Let us next consider the process 
 $ e_{\chic R} (k_1) \bar{\nu}_{\mu} (p_1) 
\to e_{\chic R} (k_2) \bar{\nu}_{\mu} (p_2) $, i.e.
the same process as before with 
$\nu_{\mu} \to \bar{\nu}_{\mu}$, and identical
kinematics.
For the tree-level contribution 
${\cal M}_{\bar{\nu}_{\mu} e_{R}}^{(0)}$ we have
\be
{\cal M}_{\bar{\nu}_{\mu} e_{R}}^{(0)} = 
 A_{(a)} \bar{\cal C}_{\chic L}
\ee
with
\be 
\bar{\cal C}_{\chic L} = [\bar{u}_{\chic R}(k_2)\gamma_{\mu} 
u_{\chic R}(k_1)]
[\bar{v}(p_1)\gamma_{\mu} P_{\chic L} \, v(p_2)]
\ee
Similarly, since the 
one-loop analysis can be repeated unaltered, we have 
for the one-loop amplitude  ${\cal M}_{\bar{\nu}_{\mu} e_{R}}^{(1)}$
\be
{\cal M}_{\bar{\nu}_{\mu} e_{R}}^{(1)} = 
\bigg(\sum_n c_n A_n \bigg) \bar{\cal C}_{\chic L} 
+ \bar{B}_{\chic Z \chic Z}
\label{ax}
\ee
with 
\bea
\bar{B}_{\chic Z \chic Z} &=& 
  + {\cal K}_{\chic Z\chic Z} 
[\bar{u}_{\chic R}(k_2)\gamma_{\rho} S_e (\ell-k_1)
\gamma_{\sigma}u_{\chic R}(k_1)]
[\bar{v}(p_1)\gamma_{\rho} P_{\chic L} S_{\bar{\nu}_{\mu}} (\ell+p_1)
\gamma_{\sigma} P_{\chic L} v(p_2)]
\nonumber\\
&& - {\cal K}_{\chic Z\chic Z}
[\bar{u}_{\chic R}(k_2)\gamma_{\rho} S_e (\ell+k_2)
\gamma_{\sigma}u_{\chic R}(k_1)]
[\bar{v}(p_1)\gamma_{\sigma} P_{\chic L} S_{\bar{\nu}_{\mu}} (\ell+p_1)
\gamma_{\rho} P_{\chic L} v(p_2)] 
\label{barBZZ}
\eea

Having set up the two amplitudes, we next turn to the
specifics of the neutrino -- anti-neutrino method. 
The basic observation 
is that the tree-level amplitudes 
${\cal M}_{\nu_{\mu} e_{\chic R}}^{(0)}$ 
as well as the
part of the one-loop amplitude ${\cal M}_{\nu_{\mu} e_{\chic R}}^{(1)}$
consisting of 
the propagator and vertex corrections, i.e. 
the first term in Eq.(\ref{ax}), which too is
proportional to the tree-level amplitude, has different 
transformation properties under the replacement 
$\nu_{\mu} \to \bar{\nu}_{\mu}$ 
than the part of the
amplitude originating from the box.
In particular, 
the coupling of the 
$Z$ boson to a pair of on-shell anti-neutrinos 
may be written in terms of on-shell neutrinos
provided that
one changes the chirality projector from $P_{\chic L}$ to 
$P_{\chic R}$
and supplying a relative minus sign \cite{Sarantakos:1983bp}. 
Specifically,
when sandwiched between external states the
$\Gamma_{Z\nu\nu}$ is given by
\be
\bar{u}(p_2)\, \Gamma_{Z\nu\nu} \,  u(p_1) =  i \bigg(\frac{g_w}{2c_w}\bigg)
\bar{u}(p_2)\gamma_{\mu} P_{\chic L} u(p_1)
\label{Gnu}
\ee
whereas 
\bea
\bar{v}(p_1) \, \Gamma_{Z\bar{\nu}\bar{\nu}}\,  v(p_2) &=& {}~~~
i \bigg(\frac{g_w}{2c_w}\bigg)\bar{v}(p_1)\gamma_{\mu} 
P_{\chic L} \, v(p_2)\nonumber\\
&=& - \, 
i \bigg(\frac{g_w}{2c_w}\bigg)\bar{u}(p_2)\gamma_{\mu} P_{\chic R} \, u(p_1)
\label{Gantinu}
\eea
Notice the crucial 
relative minus sign between Eq.(\ref{Gnu}) and
 Eq.(\ref{Gantinu}).
Under the same operation
the box contributions $\bar{B}_{\chic Z \chic Z}$
of Eq.(\ref{barBZZ})
assume the form
\bea
\bar{B}_{\chic Z \chic Z}'&=& 
  - {\cal K}_{\chic Z\chic Z}[\bar{u}_{\chic R}(k_2)
\gamma_{\rho} S_e (\ell-k_1) \gamma_{\sigma}u_{\chic R}(k_1)]
[\bar{u}(p_2)\gamma^{\sigma} P_{\chic R} S_{\nu_{\mu}} (\ell+p_1)
\gamma^{\rho} P_{\chic R} u(p_1)]
\nonumber\\
&& + {\cal K}_{\chic Z\chic Z} 
[\bar{u}_{\chic R}(k_2)\gamma_{\rho} S_e (\ell+k_2)
\gamma_{\sigma}u_{\chic R}(k_1)]
[\bar{u}(p_2)\gamma^{\rho} P_{\chic R} S_{\nu_{\mu}} (\ell+p_1)
\gamma^{\sigma} P_{\chic R} u(p_1)] 
\label{barBZZPR}
\eea
To obtain the above results, 
we simply use the fact that since the quantities considered
are scalars in the spinor space 
their values coincides with that of their transposed, and 
employ subsequently 
\be
\gamma_{\mu}^{T} = - C \gamma_{\mu}  C^{-1}, \,\,\,\,\,\,\, 
\gamma_{5}^{T} = C \gamma_{5}  C^{-1}, \,\,\,\,\,\,\, 
v^{T}(p) C  = \bar{u}(p), \,\,\,\,\,\,\, 
 C^{-1} \bar{v}^{T}(p) = u(p) .
\ee
where $C$ is the charge conjugation operator.
Thus, we can rewrite ${\cal M}_{\bar{\nu}_{\mu} e_{R}}^{(0)}$
and ${\cal M}_{\bar{\nu}_{\mu} e_{R}}^{(1)}$ as follows:
\bea
{\cal M}_{\bar{\nu}_{\mu} e_{R}}^{(0)} &=& - 
c_{(a)}  A_{(a)} {\cal C}_{\chic R}
\nonumber\\
{\cal M}_{\bar{\nu}_{\mu} e_{R}}^{(1)} &=& - 
\bigg(\sum_n c_n A_n \bigg) {\cal C}_{\chic R} 
+\bar{B}_{\chic Z \chic Z}',  
\eea
with
\be 
{\cal C}_{\chic R} = [\bar{u}_{\chic R}(k_2)\gamma_{\mu} u_{\chic R}(k_1)]
[\bar{u}(p_2)\gamma^{\mu} P_{\chic R} u(p_1)]
\ee
The above relations allow for the isolation of
the box contributions 
when judicious combinations of the forward differential cross-sections
$(d\sigma_{\nu_{e}\, e_R}/dx)_{x=1}$ and  
$(d\sigma_{\bar{\nu}_{e}\, e_R}/dx)_{x=1}$ 
are formed. 
These quantities,
up to one-loop order, are given by 
\bea
\bigg(\frac{d\sigma_{\nu_{\mu} \, e_{\chic R}}}{dx}\bigg)_{x=1} &=& 
f \Bigg[{\cal M}_{\nu_{\mu} e_{R}}^{(0)} *
{\cal M}_{\nu_{\mu} e_{R}}^{(0)\dagger} + 
2 \Re e \bigg({\cal M}_{\nu_{e} e_{R}}^{(0)}* 
{\cal M}_{\nu_{\mu} e_{R}}^{(1)\dagger }
\bigg)\Bigg]_{x=1}  \nonumber\\
\bigg(\frac{d\sigma_{\bar{\nu}_{\mu} \, e_{\chic R}}}{dx}\bigg)_{x=1}
&=& f \Bigg[ 
{\cal M}_{\bar{\nu}_{\mu} e_{R}}^{(0)}*
{\cal M}_{\bar{\nu}_{\mu} e_{R}}^{(0)\dagger } + 
2 \Re e \bigg({\cal M}_{\bar{\nu}_{\mu} e_{R}}^{(0)}*
{\cal M}_{\bar{\nu}_{\mu} e_{R}}^{(1)\dagger}\bigg)\Bigg]_{x=1} 
\eea
The $*$ in the above formulas 
denotes that the trace over initial and final fermions
must be taken, and $f \equiv (32\pi s)^{-1}$
Notice that the only source of imaginary parts
are the two boxes; since the fermions are considered to be massless 
there will be always an imaginary part for $s > 0$.
All other contributions are real,
since they all originate from $t$-channel graphs
(vertices and self-energies). 
Thus
($\bar{c}_i$ denotes the complex conjugate of $c_i$)
\bea
\bigg(\frac{d\sigma_{\nu_{\mu} \, e_{\chic R}}}{dx}\bigg)_{x=1}
&=& f 
\, \Bigg[ c_{(a)}A_{(a)}
\bigg( \bar{c}_{(a)}A_{(a)} + 2 \sum_n \bar{c}_n A_n\bigg) {\cal T}_1 
+ 2\, \Re e \bigg( \bar{c}_{(a)}A_{(a)}\, {\cal K}_{\chic Z\chic Z} 
({\cal T}_2 + {\cal T}_3) \bigg) \Bigg]_{x=1}
\nonumber\\
\bigg(\frac{d\sigma_{\bar{\nu}_{\mu} \, 
e_{\chic R}}}{dx}\bigg)_{x=1} &=&
f \, \Bigg[ c_{(a)}A_{(a)}
\bigg( \bar{c}_{(a)}A_{(a)} + 2 \sum_n \bar{c}_n A_n\bigg)\bar{\cal T}_1 
- 2\, \Re e \bigg({ \bar{c}_{(a)}A_{(a)}\, \cal K}_{\chic Z\chic Z} 
(\bar{\cal T}_2 +  \bar{\cal T}_3)  
\bigg) \Bigg]_{x=1}\nonumber\\
{}&&
\eea
with 
\bea
{\cal T}_i &=& \frac{1}{2} \bigg({\cal I}_i + {\cal I}_i^{5}\bigg)_{x=1}\, , 
\,\,\,\,\,\, i=1,2,3   \nonumber\\ 
\bar{\cal T}_i &=& \frac{1}{2} 
\bigg(\bar{\cal I}_i + \bar{\cal I}_i^{5} \bigg)_{x=1}\, , 
\,\,\,\,\,\, i=1,2,3
\label{DEQ}
\eea
where 
\bea
{\cal I}_1 &=& 
\mbox{Tr}\bigg(\gamma_{\alpha} \not\! k_2 \, \gamma_{\beta}\,
 \not\! k_1 \bigg)
\mbox{Tr}\bigg(\gamma^{\alpha} \, P_{\chic L}
\not\! p_2 \, \gamma^{\beta}\, 
P_{\chic L} \, \not\! p_1 \bigg) 
\nonumber\\ 
\bar{\cal I}_1 &=& 
\mbox{Tr}\bigg(\gamma_{\alpha} \not\! k_2 \, \gamma_{\beta}\,
 \not\! k_1 \bigg)
\mbox{Tr}\bigg(\gamma^{\alpha} \, P_{\chic R}
\not\! p_2 \, \gamma^{\beta}\, 
P_{\chic R} \, \not\! p_1 \bigg) 
\nonumber\\
{\cal I}_2  &=& - 
\mbox{Tr}\bigg(\gamma_{\alpha} \not\! k_2 \, \gamma_{\rho}\,
S_e(\ell-k_1)\,\gamma_{\sigma} \not\! k_1 \bigg) 
\mbox{Tr}\bigg(\gamma^{\alpha} \, P_{\chic L} 
\not\! p_2 \, \gamma^{\rho}\, P_{\chic L}
\,S_{\nu_{\mu}}(\ell+p_1)\,\gamma^{\sigma}\, P_{\chic L} \not\! p_1  \bigg)
\nonumber\\
\bar{\cal I}_2 &=& - 
\mbox{Tr}\bigg(\gamma_{\alpha} \not\! k_2 \, \gamma_{\rho}\,
S_e(\ell-k_1)\,\gamma_{\sigma} \not\! k_1 \bigg) 
\mbox{Tr}\bigg(\gamma^{\alpha}\, P_{\chic R} 
\not\! p_2 \, \gamma^{\sigma}\, P_{\chic R}
\,S_{\nu_{\mu}}(\ell+p_1)\,\gamma^{\rho}\, P_{\chic R} \not\! p_1  \bigg)
\nonumber\\
{\cal I}_3  &=& \mbox{Tr}\bigg(\gamma_{\alpha} \not\! k_2 \, \gamma_{\rho}\,
S_e(\ell+k_2)\,\gamma_{\sigma} \not\! k_1 \bigg) 
\mbox{Tr}\bigg(\gamma^{\alpha} \, P_{\chic L}
\not\! p_2 \, \gamma^{\sigma}\, P_{\chic L}
\,S_{\nu_{\mu}}(\ell+p_1)\,\gamma^{\rho}\, P_{\chic L} \not\! p_1  \bigg)
\nonumber\\
\bar{\cal I}_3 &=& 
\mbox{Tr}\bigg(\gamma_{\alpha} \not\! k_2 \, \gamma_{\rho}\,
S_e(\ell+k_2)\,\gamma_{\sigma} \not\! k_1 \bigg) 
\mbox{Tr}\bigg(\gamma^{\alpha} \, P_{\chic R} 
\not\! p_2 \, \gamma^{\rho}\, P_{\chic R}
\,S_{\nu_{\mu}}(\ell+p_1)\,\gamma^{\sigma}\, P_{\chic R} \not\! p_1  \bigg)
\label{Traces}
\eea
and the 
 ${\cal I}_i^{5}$ and $\bar{\cal I}_i^{5}$
are obtained from 
${\cal I}_i$ and $\bar{\cal I}_i$, respectively,
by multiplying 
the string of $\gamma$ matrices appearing in the
first trace on the right-hand sides of 
Eq.(\ref{Traces})
by a matrix $\gamma_5$. Keeping in mind that all traces are 
to be evaluated at $x=1$, 
it is straightforward to establish that
\bea
{\cal I}_i &=& \bar{\cal I}_i\, , \,\,\,\,\,\, i=1,2,3
\nonumber\\ 
{\cal I}_i^{5} &=& \bar{\cal I}_i^{5}\, , \,\,\,\,\,\, i=1,2,3
\label{LEQ}
\eea
(in fact,  
${\cal I}_1^{5} = \bar{\cal I}_1^{5} = 0$),
and therefore,  
\be
{\cal T}_i = \bar{\cal T}_i\, , \,\,\,\,\,\, i=1,2,3 . 
\label{TEQ}
\ee
In proving Eq.(\ref{LEQ}) 
we resort to the
usual properties of the Dirac $\gamma$ matrices; 
in particular, the following 
identity may be useful:
\be
\gamma_{\rho} \gamma_{\alpha} \gamma_{\sigma} =
g_{\rho\alpha}\gamma_{\sigma} + g_{\alpha\sigma}\gamma_{\rho}
- g_{\rho\sigma}\gamma_{\alpha}
+ i \epsilon_{\mu\rho\alpha\sigma}\gamma^{\mu}\gamma_{5}
\ee
We emphasize 
that the validity of the above equalities depends crucially
on the particular kinematic relations of Eq.(\ref{KINREL}), which are
themselves a direct consequence of the special forward 
limit of $t=0$ we consider.

Thus, one arrives at 
\bea
\sigma^{(-)}_{\nu_{\mu} \,e_{\chic R}} \equiv 
\bigg(\frac{d\sigma_{\nu_{\mu} \, e_{\chic R}}}{dx}\bigg)_{x=1}
- \, \bigg(\frac{d\sigma_{\bar{\nu}_{\mu} \, 
e_{\chic R}}}{dx}\bigg)_{x=1}
&=& 4\, f \,   
\Re e \bigg(\bar{c}_{(a)} A_{(a)}\, {\cal K}_{\chic Z\chic Z} 
({\cal T}_2 + {\cal T}_3) \bigg)_{x=1}
\nonumber\\
\sigma^{(+)}_{\nu_{\mu}\, e_{\chic R}} \equiv  
\bigg(\frac{d\sigma_{\nu_{\mu} \, e_{\chic R}}}{dx}\bigg)_{x=1}
+ \, \bigg(\frac{d\sigma_{\bar{\nu}_{\mu} \, e_{\chic R}}}{dx}\bigg)_{x=1}
&=& 2\, f \, 
c_{(a)}A_{(a)} \, \bigg( \bar{c}_{(a)}A_{(a)} + 
2 \sum_n \bar{c}_n A_n\bigg)
{\cal T}_1 |_{x=1}
\nonumber\\
{}&&
\label{JJJ}
\eea
Evidently 
$\sigma^{(-)}_{\nu_{\mu} \,e_{\chic R}}$ isolates the box contributions,
whereas  $\sigma^{(+)}_{{\nu}_{\mu}\, e_{\chic R}}$ 
contains only self-energy
corrections and the NCR. In particular, using that
${\cal T}_1 |_{x=1} = 4 s^2$, we obtain 
\be
\sigma^{(+)}_{\nu_{\mu}\, e_{\chic R}}
=  
\bigg(\frac{s}{4 \pi}\bigg) 
c_{(a)}A_{(a)} 
\bigg( \bar{c}_{(a)}A_{(a)} + 2 \sum_n \bar{c}_n A_n \bigg) 
\,, \,\,\,\,\,\, n= (b),(c),(d)
\label{s1}
\ee

\subsection{Unpolarized electrons}

In the neutrino -- anti-neutrino method described above 
we have used right-handedly polarized electrons, in order to
eliminate  the box graph of Fig.1g containing 
two $W$-bosons.
It turns out that, with minor modifications, this 
method may also be applied to the case of
unpolarized target fermions. 
Consider for concreteness the forward 
differential cross-sections
$(d\sigma_{e \nu_{\mu}}/dx)_{x=1}$ and
$(d\sigma_{e \bar{\nu}_{\mu}}/dx)_{x=1}$ 
corresponding to the unpolarized 
processes 
$ e \nu_{\mu} \to e \nu_{\mu}$ and  
$ e \bar{\nu}_{\mu}\to e \bar{\nu}_{\mu}$, respectively.
Now the $WW$ box graph of Fig.1g, and the one obtained from it
by letting $\nu_{\mu} \to \bar{\nu}_{\mu}$,
are present; we will denote them
by $B_{\chic W \chic W}$ and $\bar{B}_{\chic W \chic W}$,
respectively. As we will see in a moment the presence of these
graphs does not pose any problem to the generalization of the
neutrino--anti-neutrino method. 

To begin with, it is 
straightforward to verify that
all the conclusions
which were reached in the previous sub-section 
regarding the behavior of the 
Feynman graphs appearing in the polarized case 
persist in the presence of unpolarized 
electrons. Indeed, the $WW$ box aside, 
the only modification is produced by the fact that 
now the elementary vertex describing the coupling between
the electrons and the $Z$, shown in Eq.(\ref{GenVer}), 
contains also an axial part. 
This fact is however of 
no consequence for the applicability of the method, since the
only modification that the axial part will produce 
is that now the traces 
corresponding to Eq.(\ref{DEQ}) will be 
a different linear combination of the form 
\bea
{\cal T}_i &=& c_1 \,{\cal I}_i + c_2 \,{\cal I}_i^{5} \, , 
\,\,\,\,\,\, i=1,2,3   \nonumber\\ 
\bar{\cal T}_i &=& c_1 \,\bar{\cal I}_i + c_2 \,\bar{\cal I}_i^{5} \, , 
\,\,\,\,\,\, i=1,2,3
\label{DEQM}
\eea
The precise values of the coefficients $c_1$ and $c_2$ may be 
easily worked out, but are immaterial for our arguments,   
due to the validity of  Eq.(\ref{LEQ}).  

Turning to the $WW$ box, we have for the neutrino case
\be
B_{\chic W \chic W} 
 - {\cal K}_{\chic W \chic W} 
[\bar{u}(k_2)\gamma_{\rho} P_{\chic L} S_{\nu_e} (\ell-k_1)
\gamma_{\sigma} P_{\chic L} u(k_1)]
[\bar{u}(p_2)\gamma_{\rho} P_{\chic L} S_{\mu} (\ell+p_1)
\gamma_{\sigma} P_{\chic L} u(p_1)]
\ee 
whereas for the anti-neutrino
\be 
\bar{B}_{\chic W \chic W} = + 
{\cal K}_{\chic W \chic W} 
[\bar{u}(k_2)\gamma_{\rho} P_{\chic L} S_{\nu_e} (\ell-k_1)
\gamma_{\sigma} P_{\chic L} u(k_1)]
[\bar{v}(p_1)\gamma_{\rho} P_{\chic L} S_{\mu} (\ell+p_1)
\gamma_{\sigma} P_{\chic L} v(p_2)]
\ee
with
\be
{\cal K}_{\chic W\chic W} 
\equiv \frac{g_w^4}{4}
\int \frac{d^4\ell}{(2\pi)^4}
D_{\chic W} (\ell)  D_{\chic W} (\ell+p_1-p_2) 
\ee
and $D_{\chic W}(k)= (k^2 -M_{\chic W}^2)^{-1}$ .
Then, under the aforementioned transformation 
$\bar{B}_{\chic W \chic W}$ becomes
\be 
\bar{B}_{\chic W \chic W}' = - 
{\cal K}_{\chic W \chic W} 
[\bar{u}(k_2)\gamma_{\rho} P_{\chic L} S_{\nu_e} (\ell-k_1)
\gamma_{\sigma} P_{\chic L} u(k_1)]
[\bar{u}(p_2)\gamma_{\sigma} P_{\chic R} S_{\mu} (\ell+p_1)
\gamma_{\rho} P_{\chic R} u(p_1)]\,.
\ee
Therefore, the $WW$ boxes may be isolated 
into $\sigma^{(+)}_{\nu_{\mu} e} \equiv
(d\sigma_{\nu_{\mu} e}/dx)_{x=1} 
- (d\sigma_{\bar{\nu}_{\mu} e}/dx )_{x=1}$ 
exactly as happened with the 
 $ZZ$ boxes, provided that the relevant traces
\bea
{\cal T}_4  &=&  
\mbox{Tr}\bigg(\gamma_{\alpha}\, (a_{e}+b_{e}\gamma_{5})\,
\not\! k_2 \, \gamma_{\rho}\, P_{\chic L}\,
S_{\nu_e}(\ell-k_1)\,\gamma_{\sigma} \, P_{\chic L}\, \not\! k_1 \bigg) 
\mbox{Tr}\bigg(\gamma^{\alpha} \, P_{\chic L} 
\not\! p_2 \, \gamma^{\rho}\, P_{\chic L}
\,S_{\mu}(\ell+p_1)\,\gamma^{\sigma}\, P_{\chic L} \not\! p_1  \bigg)
\nonumber\\
\bar{\cal T}_4 &=&  
\mbox{Tr}\bigg(\gamma_{\alpha} \, (a_{e}+b_{e}\gamma_{5})\,  
\not\! k_2 \, \gamma_{\rho}\,  P_{\chic L}\,
S_{\nu_e}(\ell-k_1)\,\gamma_{\sigma} \,  P_{\chic L}\,\not\! k_1 \bigg) 
\mbox{Tr}\bigg(\gamma^{\alpha}\, P_{\chic R} 
\not\! p_2 \, \gamma^{\sigma}\, P_{\chic R}
\,S_{\mu}(\ell+p_1)\,\gamma^{\rho}\, P_{\chic R} \not\! p_1  \bigg)
\nonumber
\eea
appearing in the
two cross-sections are equal; 
this is indeed so, as an immediate consequence of
Eq.(\ref{LEQ}).
Thus, we can arrive at the analogue of  the
second relation in Eq.(\ref{JJJ}), which
now will be given by 
\bea 
\sigma^{(+)}_{\nu_{\mu} e}  &\equiv& 
\bigg(\frac{d\sigma_{\nu_{\mu} e}}{dx}\bigg)_{x=1} 
+ \, \bigg(\frac{d\sigma_{\bar{\nu}_{\mu} e}}{dx}\bigg)_{x=1} 
\nonumber\\
&=& 2\, f \, 
C_{(a)}^{e} A_{(a)} \, \Bigg[ \bigg(\bar{C}_{(a)}^{e} A_{(a)} +
2 \bar{C}_{(b)}^{e} A_{(b)}\bigg) {\cal T}_{\chic A} +
2 \bigg(\bar{C}_{(c)}^{e} A_{(c)} + \bar{C}_{(d)}^{e} A_{(d)}\bigg)
{\cal T}_{\chic B} \Bigg]_{x=1}
\label{JJJ2}
\eea
with
$C_{(a)}^{e} = C_{(b)}^{e} = i g^2_w / 2 c^2_w$, 
${C}_{(c)}^{e} = - \,i e (g_w/2 c_w)$, 
${C}_{(d)}^{e} =  -\, 2 i e^2$
and
\bea
{\cal T}_{\chic A}|_{x=1} &=&  
\Bigg[\mbox{Tr}\bigg(\gamma_{\alpha}\,(a_{e} + b_{e} \gamma_5)\,
\not\! k_2 \, \gamma_{\beta}\,(a_{e} + b_{e} \gamma_5)\,
 \not\! k_1 \bigg)
\mbox{Tr}\bigg(\gamma^{\alpha} \, P_{\chic L}
\not\! p_2 \, \gamma^{\beta}\, 
P_{\chic L} \, \not\! p_1 \bigg)\Bigg]_{x=1} \nonumber\\
&=& 8 (a_{e}^2 + b_{e}^2) s^2 
\nonumber\\
{\cal T}_{\chic B}|_{x=1} &=&
\Bigg[\mbox{Tr}\bigg(\gamma_{\alpha}\,(a_{e} + b_{e} \gamma_5)\,
\not\! k_2 \, \gamma_{\beta}\,\not\! k_1 \bigg)
\mbox{Tr}\bigg(\gamma^{\alpha} \, P_{\chic L}
\not\! p_2 \, \gamma^{\beta}\, 
P_{\chic L} \, \not\! p_1 \bigg)\Bigg]_{x=1}
=  8\, a_{e}\, s^2 
\label{JTR}
\eea
Thus,
from 
Eq.(\ref{JJJ2})
we finally obtain for $\sigma^{(+)}_{\nu_{\mu} e}$
\be
\sigma^{(+)}_{\nu_{\mu} e} = \bigg(\frac{s}{2\pi}\bigg)
C_{(a)}^{e} A_{(a)} \, \Bigg[ \bigg(\bar{C}_{(a)}^{e} A_{(a)} +
2 \bar{C}_{(b)}^{e} A_{(b)}\bigg) (a_{e}^2 + b_{e}^2) +
2 a_{e} \bigg(\bar{C}_{(c)}^{e} A_{(c)} + \bar{C}_{(d)}^{e} A_{(d)}\bigg)
 \Bigg]
\label{JJJB}
\ee
The actual values of $a_{e}$ and $b_{e}$ are obtained
from  Eq.(\ref{GenVer})
by setting $Q_{e} =-1$ and 
$T^{e}_z = -\frac{1}{2}$, i.e. 
$a_{e}= \frac{1}{4} - s^2_w $ and 
$b_{e}= - \frac{1}{4}$. 

\subsection{Neutrino scattering}

It is easy to see that the same method used to arrive at 
Eq.(\ref{JJJ2}) may be employed in order to isolate 
{\it directly}
the universal,  $Z$-mediated contribution 
from
experiments involving
{\it only} neutrinos and anti-neutrinos (see also section 
\ref{sec:EXT}) 
In particular
we will apply the neutrino -- anti-neutrino method
to the two processes 
$\nu_{e} \nu_{\mu}  \to \nu_{e} \nu_{\mu} $ and  
$ \nu_{e} \bar{\nu}_{\mu} \to  \nu_{e} \bar{\nu}_{\mu} $. 
The corresponding 
graphs can be obtained from those of Fig.1 (except (1c) and (1d)), 
by replacing in the first case the external electrons by
neutrinos, and in the second case by 
replacing the external  electrons by
neutrinos and the external neutrinos by anti-neutrinos.
Regarding these processes, the following points are important: 
First, the vertex corrections vanish as before in the
kinematic limit of $t=0$.
Second, the only universal combinations is precisely that
mediated by the $Z$. Third, 
one can eliminate all box contributions
if one considers the
quantity
$\sigma^{(+)}_{\nu_{\mu} \nu_{e}} \equiv 
(d\sigma_{\nu_{\mu} \nu_{e} }/dx)_{x=1} + 
(d\sigma_{\bar{\nu}_{\mu} \nu_{e} }/dx)_{x=1}$ ;
this happens exactly as before, resorting only to 
results presented in Section II, such as 
Eq.(\ref{TEQ}). 
Thus, one can show that   
\be
\sigma^{(+)}_{\nu_{\mu} \nu_{e}} =  \bigg(\frac{s}{16 \pi}\bigg) \, 
\bigg(\frac{ g^4_w}{c_w^4}\bigg) \,
 A_{(a)} \, \Bigg[A_{(a)} + 2 A_{(b)}\Bigg]
\label{s3}
\ee

\subsection{General case}

It is now straightforward to generalize the above analysis 
for the general case of a neutrino $\nu_{i}$ scattering off 
a target fermion $f$, which may be 
a charged lepton 
other than the iso-doublet partner of $\nu_{i}$, 
a quark, or a neutrino, all of which may have any polarization.
The general formula reads 
 \be
\sigma^{(+)}_{\nu_{i} f} = \bigg(\frac{s}{2\pi}\bigg)
C_{(a)}^{f} A_{(a)} \, \Bigg[ \bigg(\bar{C}_{(a)}^{f} A_{(a)} +
2 \bar{C}_{(b)}^{f} A_{(b)}\bigg) (a_{f}^2 + b_{f}^2) +
2 a_{f} \bigg(\bar{C}_{(c)}^{f} A_{(c)} + \bar{C}_{(d)}^{f} A_{(d)}\bigg)
 \Bigg]
\label{JJJN}
\ee
with 
$C_{(a)}^{f} = C_{(b)}^{f} = i g^2_w / 2 c^2_w$, 
${C}_{(c)}^{f} =  \,i e Q_f (g_w/2 c_w)$, 
${C}_{(d)}^{f} =  \, 2 i e^2 Q_f$. 

Evidently, $\sigma^{(+)}_{\nu \nu}$ of  Eq.(\ref{s3})
may be obtained from 
Eq.(\ref{JJJN}) by setting $Q_f=0$, i.e. 
by setting $C_{(c)}^{\nu}= C_{(d)}^{\nu} = 0$, and
$a_{\nu} = - \frac{1}{2}$,   
$b_{\nu} = \frac{1}{2}$. Similarly, $\sigma^{(+)}_{\nu_{\mu} e}$
of Eq.(\ref{JJJB})
is obtained by setting $Q_f=Q_e=-1$. Finally, to 
recover the right-handedly polarized case 
$\sigma^{(+)}_{\nu_{\mu}\, e_{\chic R}}$ of 
Eq.(\ref{s1}), we must use in Eq.(\ref{JJJN}) the effective 
$a_{e_{\chic R}}$ and $b_{e_{\chic R}}$ obtained when 
casting the expression for $\Gamma_{Z {f_R} \bar{f_R}}^{\mu}$
of Eq.(\ref{Verer}) into that given in the 
last line of Eq.(\ref{GenVer});
in particular,  
$a_{f_{\chic R}} = b_{f_{\chic R}} = Q_f (s_w^2/2) $ 
and $a_{e_{\chic R}} = b_{e_{\chic R}} = - s_w^2/2$. 

Summarizing the results of this section until this point, 
we have demonstrated that  
the appropriate combination of neutrino and anti-neutrino
amplitudes has allowed us to discard the remaining process-dependent 
contributions related to the boxes. We conclude this section by 
providing a direct way of measuring the {\it difference} 
between the NCR of two neutrinos of different flavor.

\subsection{Measuring the difference  
$\big <r^2_{\nu_i}\,  \big>$ - $\big <r^2_{\nu_j}\,  \big>$. }

Using the results of this section 
one can extract the values 
of the observables $r_{ij}$ defined as the difference of the
NCR of different neutrino flavours, i.e.:
\bea
r_{ij}  = \big< r^2_{\nu_i}\, \big>
 - \big< r^2_{\nu_j}\, \big> 
\,\, , \,\,\,\,\,\,i,j=e,\mu,\tau, \,\,\,\,\,\, i\neq j . 
\eea
Clearly, $r_{ij} = - r_{ji}$. Notice 
that there are only  
two such independent observables, since one can write any of the
three $r_{ij}$ as a linear combination
of the other two, e.g. 
$r_{\tau\mu} = r_{\tau e} - r_{\mu e}$.

To see that in detail, we turn to Eq.(\ref{JJJN}),
and  
let us consider the difference  
between the $\sigma^{(+)}_{\nu_{\mu}\, e}$ 
$\sigma^{(+)}_{\nu_{\tau}\, e}$; clearly,
the only difference between the two 
is the replacement of 
$\widehat{F}_{\nu_{\mu}}(0)$ by $\widehat{F}_{\nu_{\tau}}(0)$, 
and correspondingly  
 $\big< r^2_{\nu_{\mu}}\, \big>$ by 
$\big< r^2_{\nu_{\mu}}\, \big>$,  in the $A_{(d)}$
of Eq.(\ref{THEA}), whereas all remaining terms are 
the same. 
Thus, when forming the difference the (ultraviolet divergent) 
universal parts cancel, and we are left with 
\be
\sigma^{(+)}_{\nu_{\mu}\, e } - \sigma^{(+)}_{\nu_{\tau}\, e } 
= \lambda\, (1- 4 s_w^2)\, r_{\mu\tau}
\label{RIJ}
\ee
where  $\lambda \equiv (2\sqrt{2}/3)\, s \,\alpha \,G_{\chic F}$, 
and $\alpha = e^2/4\pi$ is the fine-structure constant.
Clearly, to obtain the quantity $r_{e\tau }$ one must use
muons as target fermions, and in order to obtain
$r_{e\mu}$ one must use taus.

We note that, a priori, the difference in the 
forward amplitudes ${\cal M}_{\nu_{\mu} e} - {\cal M}_{\nu_{\tau} e}$  
would contribute to a difference for the neutrino index of refraction 
\cite{Botella:1986wy} in electron matter; this difference vanishes, 
however, for ordinary matter due to its neutrality.

\section{Renormalization-group analysis}\label{sec:RGI}

In the previous sections we have demonstrated that, in the 
kinematic limit of interest, the appropriate combination 
of physical cross-section, $\sigma^{(+)}_{\nu_{i} f}$, can be 
finally expressed in terms of the gauge-invariant 
and universal one-loop  PT self-energies 
${\widehat{\Pi}}^{ \chic{A}  {\chic  Z}}$ and   
${\widehat{\Sigma}}^{ \chic{Z}  {\chic  Z}}$, and the 
flavour-dependent $\big< r^2_{\nu_{\mu}}\, \big>$.
Contrary to the NCR which is ultra-violet finite, the self-energies
contributing to  Eq.(\ref{JJJN}) must undergo renormalization. 
In  this section  we will  show  how the aforementioned 
self-energies organize
themselves into  renormalization-group invariant (RGI)  quantities. An
immediate by-product of the analysis presented in this section is that
the  {\it renormalized}\,   ${\widehat{\Pi}}^{ \chic{A}  {\chic  Z}} (0)$
{\it cannot}  form part of the NCR,  because 
it fails  to form a
RGI quantity on its own. The general framework presented in this
section has been established in 
\cite{Denner:1994xt,Papavassiliou:1996fn,Papavassiliou:1997pb};  
here we will adopt the notation and philosophy developed in 
\cite{Papavassiliou:1997pb},
focusing mainly on the aspects relevant for the problem 
at hand.  

The quantity that serves as the field-theoretic prototype 
for the construction presented in this section  
is the effective charge of QED,  which is  
a RGI
and, at the same time, universal, i.e. process-independent, quantity.
Its construction, which 
constitutes text-book material \cite{Itzykson:rh}, proceeds as follows:
One begins by considering  
the unrenormalized photon self-energy   is $\Pi^0_{\mu\nu}(q) =  (q^2
g_{\mu\nu}  -  q_{\mu}q_{\nu})\Pi^0(q^2)$,  where  $\Pi^0 (q^2)$ is  a
gauge-independent function to all  orders in perturbation theory.  After
performing the  standard  Dyson   summation,  we obtain   the  dressed
photon propagator between conserved external currents
$\Delta^0_{\mu\nu}(q)\ = 
(g_{\mu\nu}/q^2) [1+\Pi^0(q^2)]^{-1}$ \, .
The   above   quantity   is   universal,   in   the   sense  that   is
process independent. The renormalization procedure introduces 
the standard relations  between  renormalized    and unrenormalized
parameters:  
$e = Z_{e}^{-1} e^0 = Z_f Z_A^{1/2} Z_1^{-1} e^0$
and 
$1+\Pi (q^2)= Z_A [1+\Pi^0  (q^2)]$, 
where $Z_A$ and $Z_f$ are the wave-function renormalization 
constants of the
photon and fermion, respectively, and $Z_1$ the vertex renormalization, 
and $Z_{e}$ is the charge renormalization constant
The  Abelian gauge
symmetry  of the   theory   gives  rise   to  the    fundamental 
Ward identity 
$q^{\mu}\Gamma^0_{\mu}(p,p+q)= S_0^{-1}(p+q)-S_0^{-1}(p)$, where
$\Gamma^0_{\mu}$ and  $S_0 (k)$ are the  unrenormalized one-loop 
photon-electron vertex and electron propagator, respectively.
The requirement  that     the   renormalized vertex       $\Gamma_{\mu}  =
Z_{1}\Gamma^0_{\mu}$   and  the  renormalized  self-energy   $S  =
Z_{f}^{-1} S_0$ satisfy the same identity imposes the equality
$Z_{1}=Z_{f}$, from which immediately follows that 
$Z_{e}\ =\ Z_{A}^{-1/2}$. Given these relations between 
the renormalization constants,
we can now form the following RGI combination:
\begin{equation}
  \label{alphaqed}
\bar{R}_{\mu\nu}(q^2)\ =
\ \frac{(e^0)^2}{4\pi}\, \Delta^0_{\mu\nu}(q)
\ =\ \frac{e^2}{4\pi}\, \Delta_{\mu\nu}(q)\, .
\end{equation}   
From $\bar{R}_{\mu\nu}(q^2)$, after pulling out a 
the trivial kinematic factor $(1/q^2)$, one may define 
the usual QED effective charge  $\bar{\alpha} (q^2)$. 
This effective charge 
has a non-trivial dependence on the 
masses $m_i$  of  the  particles appearing  in the 
vacuum polarization loop,  
which   allows its   reconstruction  from  physical amplitudes, 
by resorting to the optical theorem and analyticity, i.e. dispersion 
relations.  
For  $q^2\gg  m^2_i$,
$ \bar{\alpha} (q^2)$ 
coincides  with  the  one-loop  running coupling  of  the
theory, i.e. the solution of the standard renormalization-group 
equation. 

In non-Abelian gauge theories  the crucial equality $Z_1=Z_f$ does not
hold in  general, because the Ward identities are replaced by 
the more complicated Slavnov-Taylor identities 
\cite{Slavnov:1972fg,Taylor:1971ff}, involving ghost Green's functions.  
Furthermore, in contrast to   the photon  case, the
vacuum  polarization of the gauge bosons 
depends on  the gauge-fixing parameter, already  at one-loop
order.  These  facts make the  non-Abelian  generalization  of the QED
concept of  the  effective  charge  more complicated 
\cite{Cornwall:1982zr,Papavassiliou:1996fn,Papavassiliou:1997pb, 
Watson:1996fg}.  
The PT   rearrangement  of   physical amplitudes  gives   rise   to  a
gauge-independent effective self-energy,  and restores at the same
time QED-like Ward identities. 
To see  this in detail for the case at hand, 
we start  by listing the relations between the
bare and renormalized parameters for the electroweak sector of the SM.
We indicate all  (bare) unrenormalized quantities with the superscript
`0'. For the masses we have
\begin{equation}
  \label{massren}
(M^0_W)^2 = M_W^2\ +\ \delta M_W^2\, , \qquad
(M^0_Z)^2\ =\ M_Z^2\ +\ \delta M_Z^2\, , 
\end{equation}
The wave-function renormalizations for the neutral sector are defined as
\be
\left( \begin{array}{c} Z_{0} \\ A_{0}
\end{array} \right)\ =
\left( \begin{array}{cc}
\widehat{Z}_{\chic{Z} \chic{Z}} & 
\widehat{Z}_{ \chic{Z} \chic{A}} \\
\widehat{Z}_{\chic{A} \chic{Z}}        
& \widehat{Z}_{\chic{A}\chic{A}}
\end{array} \right)\,
\left( \begin{array}{c} Z \\ A
\end{array} \right)
\ = 
\left( \begin{array}{cc}
1+\frac{1}{2}\delta \widehat{Z}_{\chic{Z} \chic{Z}} & 
\frac{1}{2}\delta \widehat{Z}_{ \chic{Z} \chic{A}} \\
\frac{1}{2}\delta \widehat{Z}_{\chic{A} \chic{Z}}        
&1+\frac{1}{2}\delta \widehat{Z}_{\chic{A}\chic{A}}
\end{array} \right)\,
\left( \begin{array}{c} Z \\ A
\end{array} \right)
\label{GammaNCINV}
\ee

In addition, the coupling renormalization constants are defined by
\be
e_0 = \widehat{Z}_{e}\, e\, = 
(1 + \delta \widehat{Z}_{e})\, e\,,\,\,\,\,\,\,\,
g^0_w = \widehat{Z}_{g_w}\, g_w\, = 
(1 + \delta \widehat{Z}_{g_w})\,g_w \,,\,\,\,\,\,\,\,
c^0_w\ =\ \widehat{Z}_{c_{w}}c_{w}\, ,
\ee

with
\begin{equation}
\widehat{Z}_{c_w}\ =\ \Big(\, 1\, +\, \frac{\delta M_W^2}{M_W^2}\, \Big)^{1/2}
\Big(\, 1\, +\, \frac{\delta M_Z^2}{M_Z^2}\, \Big)^{-1/2}\, .
\end{equation}
If we expand $\widehat{Z}_{c_{w}}$ perturbatively, we have
that
$\widehat{Z}_{c_{w}}\ =\ 1+ \frac{1}{2} (\delta c^2_w / c^2_w) + \dots\,,$
with
\begin{equation}
\frac{\delta c^2_w}{c^2_w}\ =\ \frac{\delta M_W^2}{M_W^2}-
\frac{\delta M_Z^2}{M_Z^2}\, ,
\end{equation}
which is the usual one-loop result.  

Imposing the requirement that the  PT Green's functions should respect
the  same  WI's  before and   after renormalization  we  arrive at the
following relations:
\be
\widehat{Z}_{\chic{A}\chic{A}} =  \widehat{Z}_{e}^{-2} \, ,\,\,\,\,\,\,\,\,\,\,
\widehat{Z}_{\chic{Z}\chic{Z}} = {\widehat{Z}}_{g_w}^{-2}
\widehat{Z}_{c_{w}}^2 \, ,
\label{W3}
\ee
or, equivalently, at the level of the counter-terms
\begin{eqnarray}
\delta\widehat{Z}_{\chic{A}\chic{A}} &=& -2 \delta \widehat{Z}_{e}\, ,
\nonumber\\
\delta\widehat{Z}_{\chic{Z}\chic{Z}} &=& -2 \delta \widehat{Z}_{e}
- \frac{c^2_w - s^2_w}{s^2_w} \bigg(\frac{\delta c^2_w}{c^2_w}\bigg)\, ,
\nonumber\\ 
\delta \widehat{Z}_{\chic{A}\chic{Z} } &=& 
2 \frac{c_w}{s_w} \bigg(\frac{\delta c^2_w}{c^2_w}\bigg)\, ,
\nonumber\\
\delta \widehat{Z}_{\chic{Z}\chic{A} } &=& 0 \, .
\end{eqnarray}

The corresponding propagators relevant for the neutral sector
may be obtained by inverting the matrix $\widehat{L}$, whose 
entries are the PT self-energies, i.e 
\be
\widehat{L} =
\left( \begin{array}{cc}
q^2 + \widehat{\Sigma}_{\chic{A} \chic{A}} & 
\widehat{\Sigma}_{\chic{A}\chic{Z}} \\
\widehat{\Sigma}_{\chic{A} \chic{Z} }         
& q^{2} - M_{Z}^{2} + \widehat{\Sigma}_{\chic{Z}\chic{Z} }
\end{array} \right)\,.
\label{GammaNC}
\ee
Casting the inverse in the form
\be
\widehat{L}^{-1} =
\left( \begin{array}{cc}
\widehat{\Delta}_{\chic{A} \chic{A}}(q^2) & 
\widehat{\Delta}_{ \chic{A} \chic{Z}}(q^2) \\
\widehat{\Delta}_{\chic{A} \chic{Z}}(q^2)        
& \widehat{\Delta}_{\chic{Z} \chic{Z}(q^2) }
\end{array} \right)\,.
\label{GammaNCI}
\ee
one finds that 
\bea
\widehat{\Delta}_{\chic{A} \chic{A}}(q^2) &=& 
\frac{- [q^{2} -M_{Z}^{2} + 
\widehat{\Sigma}_{\chic{Z}\chic{Z}}(q^2)]}
{\widehat{\Sigma}_{\chic{A}\chic{Z}}^{2}(q^2) -
[q^{2} -M_{Z}^{2} + \widehat{\Sigma}_{\chic{Z}\chic{Z} }(q^2)]
[q^{2} + \widehat{\Sigma}_{\chic{A}\chic{A}}(q^2)]}
\,,\nonumber\\
\widehat{\Delta}_{\chic{Z}\chic{Z}}(q^2)      &=&
\frac{- [q^{2} + \widehat{\Sigma}_{\chic{A}\chic{A}}(q^2)]}
{\widehat{\Sigma}_{\chic{A}\chic{Z}}^{2}(q^2) -
[q^{2} -M_{Z}^{2} + \widehat{\Sigma}_{\chic{Z}\chic{Z} }(q^2)]
[q^{2} + \widehat{\Sigma}_{\chic{A}\chic{A}}(q^2)]}
\,,\nonumber\\
\widehat{\Delta}_{\chic{A} \chic{Z}}(q^2) &=& 
\frac{\widehat{\Sigma}_{\chic{A}\chic{Z}}(q^2)}
{\widehat{\Sigma}_{\chic{A}\chic{Z}}^{2}(q^2) -
[q^{2} -M_{Z}^{2} + \widehat{\Sigma}_{\chic{Z}\chic{Z} }(q^2)]
[q^{2} + \widehat{\Sigma}_{\chic{A}\chic{A}}(q^2)]}\,,
\eea
The above expressions at one-loop reduce to 
\bea
\widehat{\Delta}_{\chic{A} \chic{A}} (q^2) &=& 
\frac{1}
{q^{2} + \widehat{\Sigma}_{\chic{A}\chic{A}}(q^2)}
\,,\nonumber\\
\widehat{\Delta}_{\chic{Z}\chic{Z}}(q^2) &=&
\frac{1}
{q^{2} -M_{Z}^{2} + \widehat{\Sigma}_{\chic{Z}\chic{Z}}(q^2)}
\,,\nonumber\\
\widehat{\Delta}_{\chic{A} \chic{Z}}(q^2) &=& 
\frac{-\widehat{\Sigma}_{\chic{A}\chic{Z}}(q^2)}
{q^{2} (q^{2} -M_{Z}^{2})}\,,
\eea

The standard re-diagonalization procedure of the neutral sector  
\cite{Baulieu:ux,Kennedy:1988sn,Philippides:1995gf}
may then be followed,  
for the PT self-energies; it will finally amount to 
introducing the effective (running) weak mixing angle.
In particular,  
after the PT rearrangement, 
the propagator-like part $\widehat{\cal D}_{ff'}$ 
of the neutral current
amplitude for the interaction between fermions
with charges $Q$, $Q'$ and isospins $T^{f}_z $,  $T^{f'}_z $,  
is given
in terms of the inverse of the matrix $\widehat{L}$ by the expression
\bea
\widehat{\cal D}_{ff'}\, &=&\, 
\lefteqn{
\left( \begin{array}{cc}
e Q_{f}, & {\displaystyle \frac{g_w}{c_w}}
( s^2_w Q_{f} - T^{f}_z P_{\chic L} )
\end{array} \right)
\widehat{L}^{-1}
\left( \begin{array}{c}
e Q_{f'} \\ {\displaystyle \frac{g_w}{c_w}}
(s^2_w Q_{f'} - T^{f'}_z P_{\chic L} )
\end{array} \right) \,\,\,}\nonumber \\
&=& \left( \begin{array}{cc}
e Q_{f},  & {\displaystyle \frac{g_w}{c_w}}
\bigg[ \bar{s}^2_w(q^{2}) Q_{f} - T^{f}_z P_{\chic L}\bigg] 
\end{array} \right)
\widehat{L}^{-1}_{\chic{D}}
\left( \begin{array}{c}
e Q_{f'} \\ {\displaystyle \frac{g_w}{c_w}}
\bigg[ \bar{s}^2_w(q^{2})Q_{f'} - T^{f'}_z P_{\chic L}\bigg]
\end{array} \right)  
\label{born}
\eea
where
\be
\widehat{L}^{-1}_{\chic{D}} = 
\left( \begin{array}{cc}
\hat{\Delta}_{\chic{A} \chic{A}}(q^2)             & 0 \\ 
\phantom{\displaystyle \frac{e}{s_w}}0
\phantom{\displaystyle \frac{e}{s_w}}  & 
\hat{\Delta}_{\chic{Z} \chic{Z}}(q^2)
\end{array} \right)
\ee
The r.h.s.\ of this equation, where the neutral current interaction
between the fermions has been written in diagonal
(i.e.\ Born--like) form, defines the diagonal propagator functions
$\widehat{\Delta}_{\chic{A}\chic{A}}$ 
and $\widehat{\Delta}_{\chic{Z}\chic{Z}}$ and
the effective weak mixing angle $\bar{s}_w^{2}(q^2)$
\be\label{sweff}
\bar{s}_w^{2}(q^2) = ({s_w^{0}})^{2}\Biggl(1 +
\frac{c_w^{0}}{s_w^{0}}\frac{\widehat{\Sigma}_{\chic{A} \chic{Z}}^0(q^2)}
{q^{2}+\widehat{\Sigma}_{\chic{A}\chic{A}}^0(q^2)} \Biggr)
= s_w^{2}\Biggl(1 +
\frac{c_w}{s_w}\frac{\widehat{\Sigma}_{\chic{A}\chic{Z}}(q^2)}
{q^{2}+\widehat{\Sigma}_{\chic{A}\chic{A}}(q^2)} \Biggr)\,.
\ee
By virtue of the special relations of 
Eq.(\ref{W3}), $\bar{s}_w^{2}(q^2)$ constitutes a RGI quantity, i.e. it
retains the same form whether written in terms of bare or 
renormalized quantities.
 
At one-loop level, and after using Eq.(\ref{d4}),
$\bar{s}_w^{2}(q^2)$ reduces to 
\be\label{sweff1}
\bar{s}_w^{2}(q^2) = s_w^{2}\Biggl(1 - \frac{c_w}{s_w}\, 
{\widehat{\Pi}_{\chic{A}\chic{Z}}(q^2)}\Biggr)\,.
\ee
Notice that in the case where the fermion $f'$ is a neutrino 
($f'=\nu$, $Q_{f'}=0$ and $T^{f'}_z = 1/2$), Eq.(\ref{born}) assumes the form
\be
\widehat{\cal D}_{f\nu }\, = 
\left( \begin{array}{cc}
e Q_{f},  & {\displaystyle \frac{g_w}{c_w}}
\bigg[ \bar{s}^2_w(q^{2}) Q_{f} - T^{f}_z P_{\chic L}\bigg] 
\end{array} \right)
\widehat{L}^{-1}_{\chic{D}}
\left( \begin{array}{c}
0 \\ {\displaystyle - \frac{g_w}{2c_w}} P_{\chic L}
\end{array} \right)  
\label{born2}
\ee
Evidently, $\bar{s}_w^{2}(q^2)$ constitutes a universal   
modification to the effective charged-fermion vertex.
  
At this point it must be clear that if  
${\widehat{\Pi}}^{ \chic{A}  {\chic  Z}} (0)$ were to be considered 
as the ``universal'' part of the NCR, 
to be added to   
the ultraviolet-finite and flavour-dependent contribution 
coming  from the  proper vertex,  then  
the  resulting NCR would 
depend on  the subtraction point and scheme chosen
to renormalize it,
and would therefore be unphysical.    
Instead,
${\widehat{\Pi}}^{ \chic{A} {\chic Z}}  (0)$ must be combined with the
appropriate $Z$-mediated  {\it tree-level} contributions 
(which  evidently  do not  enter into  the
definition of the NCR) in order to form with them the RGI combination 
of Eq.(\ref{sweff1}),
whereas the ultraviolet-finite 
NCR will be determined from the proper vertex only.

The analogue of Eq.(\ref{alphaqed}) 
may be defined for 
the $Z$-boson propagator. 
In particular,
the  bare and renormalized PT resummed
$Z$-boson       propagators,      
$\widehat{\Delta}_{\chic{Z}\chic{Z},0}^{\mu\nu}(q)$
and $\widehat{\Delta}_{\chic{Z}\chic{Z}}^{\mu\nu}(q)$
respectively,     satisfy     the
following relation
\begin{equation}
  \label{renmult}
\widehat{\Delta}_{\chic{Z}\chic{Z}}^{0,\,\mu\nu}(q)\ =\ 
\widehat{Z}_{\chic{Z}\chic{Z}}\,
\, \widehat{\Delta}_{\chic{Z}\chic{Z}}^{\mu\nu}(q).
\end{equation}
In what follows we only consider the cofactors of $g^{\mu\nu}$, i.e.
$\widehat{\Delta}_{\chic{Z}\chic{Z}}^{0,\,\mu\nu}(q) =
\widehat{\Delta}_{\chic{Z}\chic{Z}}^{0}(q)\, g^{\mu\nu}$ and
$\widehat{\Delta}_{\chic{Z}\chic{Z}}^{\mu\nu}(q) =
\widehat{\Delta}_{\chic{Z}\chic{Z}}(q)\, g^{\mu\nu}$,   
since the longitudinal parts vanish when contracted with the
conserved external currents of massless fermions. 
The standard renormalization procedure  is to define the wave function
renormalization, 
$\widehat{Z}_{\chic{Z}\chic{Z}}$, by means of  the transverse part  of
the resummed $Z$-boson propagator:
\begin{equation}
  \label{TransRen}
\widehat{Z}_{\chic{Z}\chic{Z}}\, [\, q^2\, -\, (M^0_Z)^2\, +\,
\widehat{\Sigma}_{\chic{Z}\chic{Z}}^0(q^2)\, ]\ =\ q^2\, -\, M_Z^2\, +\,
\widehat{\Sigma}_{\chic{Z}\chic{Z}}(q^2)\, .
\end{equation}
It is then straightforward to verify  that the universal 
RGI quantity for the $Z$ boson, which constitutes a common part of all 
neutral current processes, is given by 
( we omit a factor  $g^{\mu\nu}$):
\be
  \label{RW}
\bar{R}_{\chic{Z}}^0(q^2)\, =  
\frac{1}{4 \pi} \bigg(\frac{g^0_w}{c^0_w}\bigg)^2
\widehat{\Delta}_{\chic{Z}\chic{Z}}^0(q^2) \, 
\ = \frac{1}{4 \pi} \bigg(\frac{g_w}{c_w}\bigg)^2
\widehat{\Delta}_{\chic{Z}\chic{Z}}(q^2)  
= \bar{R}_{\chic{Z}}(q^2) \, .
\ee
Notice that, if one retains only the real parts 
in the above equation,   
one may define from  $\bar{R}_{\chic{Z}}(q^2)$ a  
dimension-less quantity , corresponding to    
the effective charge 
$\bar{\alpha}_{\chic{Z}}(q^2)$ \cite{Hagiwara:1994pw}, by 
casting $\widehat{\Sigma}_{\chic{Z}\chic{Z}}(q^2)$ in the form 
$\widehat{\Sigma}_{\chic{Z}\chic{Z}}(q^2) = 
\widehat{\Sigma}_{\chic{Z}\chic{Z}}(M^2_Z) + 
( q^2\, -\, M_Z^2) \widehat{\Pi}_{\chic{Z}\chic{Z}}(q^2)$, and then  
pulling out a common factor $( q^2\, -\, M_Z^2)$; 
in that case, 
$\bar{R}_{\chic{Z}}^0(q^2) = ( q^2\, -\, M_Z^2)^{-1} 
\bar{\alpha}_{\chic{Z}}(q^2)$, with 
$\bar{\alpha}_{\chic{Z}}(q^2) = {\alpha}_{\chic Z}
[1+\widehat{\Pi}_{\chic{Z}\chic{Z}}(q^2)]^{-1}$, and  
${\alpha}_{\chic Z}\equiv  g_w^2 /4 \pi c_w^2$.
However, as has been explained in detail in \cite{Papavassiliou:1997pb}, 
whereas Eq.(\ref{RW}) remains valid in the presence of imaginary 
parts, i.e. when $\widehat{\Sigma}_{\chic{Z}\chic{Z}}(q^2)$ develops 
physical thresholds \cite{RES}, 
the above separation into a dimension-full and a dimension-less part  
is ambiguous and should be avoided. In the rest of this paper, 
even though we will only retain real parts, 
we will refrain from carrying out such a separation, 
expressing instead all results in terms of the more fundamental quantity 
$\bar{R}_{\chic{Z}}(q^2)$. 

Armed with the above results, we will proceed in the next section to 
separate the NCR from the rest of the tree-level and one-loop 
contributions, in a meaningful, manifestly RGI manner.

\section{Experimental extraction of the neutrino charge radius}
\label{sec:EXT}

In order to isolate the three basic quantities defined in the  
previous section we will consider three different 
processes containing them. In particular, we will study the 
differential cross-sections 
$\sigma^{(+)}_{\nu_{\mu} \,e_{\chic R}}$, 
$\sigma^{(+)}_{\nu_{\mu} \,e_{\chic L}}$,
and $\sigma^{(+)}_{\nu_{\mu} \,\nu_e}$, which
are expressed in 
Eq.(\ref{s1}), Eq.(\ref{JJJ2}), and Eq.(\ref{s3}), respectively, 
in terms of products of Feynman diagrams.
We will now use the analysis presented in the previous 
section in order to rewrite them  
in terms of the universal RGI quantities 
$\bar{R}(0)$ and $\bar{s}_w^{2}(0)$, defined in 
 Eq.(\ref{RW}) and Eq.(\ref{sweff1}), respectively,
together with the process-dependent and 
ultraviolet finite 
$\big< r^2_{\nu_{\mu}}\, \big>$. Then we obtain,
up to terms of order ${\cal O }(g_w^6)$ :
\bea
\sigma^{(+)}_{\nu_{\mu} \,e_{\chic R}}  &=&  
s \pi \bar{R}^2(0) \, \bar{s}_w^{4}(0) 
+ C_{e_{\chic R}} 
\big< r^2_{\nu_{\mu}}\, \big>
\label{syst1}\\
\sigma^{(+)}_{\nu_{\mu} \,e_{\chic L}} &=& 
s \pi \bar{R}^2(0)\bigg(\frac{1}{2} - \bar{s}_w^{2}(0)\bigg)^{2}   
+ C_{e_{\chic L}}
\big< r^2_{\nu_{\mu}}\, \big>  
\label{syst2}\\
\sigma^{(+)}_{\nu_{\mu} \,\nu_e} &=& 
s \pi \bar{R}^2(0)
\label{syst3}
\eea
with the coefficients $C_{e_{\chic R}}$ and 
$ C_{e_{\chic L}} $ given by 
\be
C_{e_{\chic R}} = - 2 \lambda s_w^{2} 
\,, \,\,\,\,\,\,\,\,\,
C_{e_{\chic L}} = 
\lambda (1-2 s_w^{2})
\,, 
\ee
The above system is linear in the unknown quantities 
$\bar{R}^2(0)$ and $\big< r^2_{\nu_{\mu}}\, \big>$, and 
quadratic in $\bar{s}_w^{2}(0)$. From 
Eq.(\ref{syst3}) we see that  
$\bar{R}^2(0)$ is already expressed in terms of the physical 
cross-section $\sigma^{(+)}_{\nu_{\mu} \,\nu}$.
This cross-section is physically important because it constitutes 
a fundamental ingredient for neutrino propagation
in a neutrino medium \cite{Notzold:1987ik}, and is relevant for 
astrophysical and cosmological scenarios. 
Thus, we are left with the system of 
Eq.(\ref{syst1}) and Eq.(\ref{syst2}) to determine 
$\bar{s}_w^{2}(0)$ and $\big< r^2_{\nu_{\mu}}\, \big>$.
Before proceeding to its solution 
notice that, as a consistency check,  
from Eq.(\ref{syst1}) and Eq.(\ref{syst2}), by changing   
$\nu_{\mu} \to \nu_{\tau}$, 
we may form the difference $r_{\mu\, \tau}$, which, 
after using that 
$ \sigma^{(+)}_{\nu_{i} \,e_{\chic R}} + 
\sigma^{(+)}_{\nu_{i} \,e_{\chic L}} =
\sigma^{(+)}_{\nu_{i} \,e}$, 
coincides with the expressions obtained in  
Eq.(\ref{RIJ}), as it should. 

The corresponding solutions of the system 
are given by
\bea
\bar{s}_w^{2}(0) &=&  s_w^{2} \pm 
\sqrt{\Omega_{\nu_{\mu}}}
\label{sol1}\\
\big< r^2_{\nu_{\mu}}\, \big> &=&  
\lambda^{-1}
\Bigg[\bigg(s_w^{2}-\frac{1}{4} \pm 
\sqrt{\Omega_{\nu_{\mu}}}\bigg)
\sigma^{(+)}_{\nu_{\mu} \,\nu_e}  
+ \sigma^{(+)}_{\nu_{\mu} \,e_{\chic L}} - 
\sigma^{(+)}_{\nu_{\mu} \,e_{\chic R}}
\Bigg]
\label{sol2}
\eea
where the discriminant $\Omega_{\nu_{\mu}}$ is given by
\be
\Omega_{\nu_{\mu}} = 
(1- 2 s_w^{2}) \bigg(\frac{\sigma^{(+)}_{\nu_{\mu} \,e_{\chic R}}}
{\sigma^{(+)}_{\nu_{\mu} \,\nu}} -  
\frac{1}{2} s_w^{2}  
\bigg)
+  2 s_w^{2} 
\frac{\sigma^{(+)}_{\nu_{\mu} \,e_{\chic L}}}
{\sigma^{(+)}_{\nu_{\mu} \,\nu}}
\label{disc}
\ee
We can again check the self-consistency of these 
solutions 
by forming the difference $r_{\mu\, \tau}$ using 
the expressions for 
$\big< r^2_{\nu_{\mu}}\, \big>$ and 
$\big< r^2_{\nu_{\tau}}\, \big>$ given in 
Eq.(\ref{sol2}), making the appropriate replacements 
(e.g. $\Omega_{\nu_{\mu}}\to \Omega_{\nu_{\tau}}$);
it is straightforward to verify that one arrives at 
a trivial identity (as one should), provided that 
$\sigma^{(+)}_{\nu_{\mu} \,\nu_e} = 
\sigma^{(+)}_{\nu_{\tau} \,\nu_e}$, which is automatically 
true, by virtue of Eq.(\ref{syst3}).
Finally, 
the actual sign in front of 
$\Omega_{\nu_{\mu}}$ may be chosen by requiring that 
it correctly accounts for the sign of the shift of $\bar{s}_{w}^{2}(0)$
with respect to $ s_w^{2}$ 
predicted by the theory \cite{Hagiwara:1994pw}.

To extract the experimental values of the quantities 
$\bar{R}^2(0)$, $\bar{s}_w^{2}(0)$, and $\big< r^2_{\nu_{\mu}}\, \big>$,
one must substitute in Eq.(\ref{syst3}), as well as 
in Eq.(\ref{sol1}) -- Eq.(\ref{disc}) the experimentally
measured values for the differential cross-sections 
$\sigma^{(+)}_{\nu_{\mu} \,e_{\chic R}}$, 
$\sigma^{(+)}_{\nu_{\mu} \,e_{\chic L}}$,
and $\sigma^{(+)}_{\nu_{\mu} \,\nu}$. 
Evidently, in order to solve this system one would 
have to carry out three different pairs of experiments. 

Another possibility, 
is to 
consider up- and down-quarks as
target fermions and 
combine appropriately 
the corresponding cross-sections 
$\sigma^{(+)}_{\nu_{\mu} \,u}$ and 
$\sigma^{(+)}_{\nu_{\mu} \,d}$ with the 
unpolarized electron cross-section
$\sigma^{(+)}_{\nu_{\mu} \,e}$. We omit QCD effects  
related to the external (target) quarks throughout;
as has been explained in \cite{Degrassi:1992ue}, 
the current algebra formulation of the PT guarantees
that the electroweak constructions used in this paper go through  
even in the presence of strong interaction effects.

In particular
we have 
\bea
\sigma^{(+)}_{\nu_{\mu} \,u}  &=&  
s \pi \bar{R}^2(0)
\bigg(\frac{1}{4} - \frac{2}{3} \bar{s}_w^{2}(0)
+ \frac{8}{9} \bar{s}_w^{4}(0)\bigg) 
+ C_{u} 
\big< r^2_{\nu_{\mu}}\, \big>
\nonumber\\
\sigma^{(+)}_{\nu_{\mu} \,d} &=& 
s \pi \bar{R}^2(0)
\bigg(\frac{1}{4} - \frac{1}{3} \bar{s}_w^{2}(0)
+ \frac{2}{9} \bar{s}_w^{4}(0)\bigg) 
+ C_{d}
\big< r^2_{\nu_{\mu}}\, \big>  
\nonumber\\
\sigma^{(+)}_{\nu_{\mu} \,e} &=& 
s \pi \bar{R}^2(0)
\bigg(\frac{1}{4} -  \bar{s}_w^{2}(0)
+ 2 \,\bar{s}_w^{4}(0)\bigg) 
+ C_{e}
\big< r^2_{\nu_{\mu}}\, \big>  
\label{systel}
\eea
with 
\be
C_{u} = \frac{2}{3} \bigg(1- \frac{8}{3} s_w^{2}\bigg)\,\lambda 
\,, \,\,\,\,\,\,
C_{d} = \frac{1}{3} \bigg(1- \frac{4}{3} s_w^{2}\bigg)\,\lambda 
\,,\,\,\,\,\,\,   
C_{e} = \bigg(1- 4 s_w^{2}\bigg)\,\lambda 
\,,
\ee
Defining the following two auxiliary linear combinations 
of the relevant cross-sections, 
\bea
w^{(+)}_{\nu_{\mu}} &\equiv& 4 \Bigg[ 
\sigma^{(+)}_{\nu_{\mu} \,e} + 
3 \bigg(\sigma^{(+)}_{\nu_{\mu} \,d}  - 
\sigma^{(+)}_{\nu_{\mu} \,u}\bigg)\Bigg] \,,
\nonumber\\
z^{(+)}_{\nu_{\mu}} &\equiv&  \frac{1}{2}
\bigg(9\, \sigma^{(+)}_{\nu_{\mu} \,d} - \sigma^{(+)}_{\nu_{\mu} \,e} \bigg)
\,,
\eea
the above system yields the following solutions for the three 
unknown quantities $\bar{R}^2(0)$, $\bar{s}_w^{2}(0)$, and
$\big< r^2_{\nu_{\mu}}\, \big>$ :
\bea
\bar{R}^2(0) &=&  \bigg(\frac{1}{s \pi}\bigg) w^{(+)}_{\nu_{\mu}} 
\nonumber\\
\bar{s}_w^{2}(0) &=& s_w^{2} \pm \sqrt{\Omega_{\nu_{\mu}} }
\nonumber\\
\big< r^2_{\nu_{\mu}}\, \big> &=& 
\lambda^{-1}\bigg[ z^{(+)}_{\nu_{\mu}} - c_w^2 w^{(+)}_{\nu_{\mu}}    
\pm w^{(+)}_{\nu_{\mu}}\sqrt{\Omega_{\nu_{\mu}}}\bigg]
\eea
where the discriminant $\Omega_{\nu_{\mu}}$ is now given by 
\be
\Omega_{\nu_{\mu}} = \frac{3}{8} + s_w^2 (s_w^2 -2) +
\frac{1}{2 w^{(+)}_{\nu_{\mu}}}
\bigg[ \sigma^{(+)}_{\nu_{\mu} \,e}
 - (1-4s_w^2) z^{(+)}_{\nu_{\mu}}\bigg]
\ee 
Clearly we must have that $w^{(+)}>0$ and 
$\Omega_{\nu_{\mu}} >0$ . 

Finally, we report the numerical values of the theoretical predictions
for the three basic parameters, 
$\bar{R}(0)$, $\bar{s}_w^{2}(0)$, and $\big< r^2_{\nu_e}\, \big>$

The numerical evaluation of Eq.(\ref{ncr}) for  
the three different neutrino flavors yields \cite{Bernabeu:2000hf}
\bea
\big< r^2_{\nu_e}\, \big> &=& 
4.1 \times 10^{- 33}\ \mathrm{cm^2} \nonumber\\
\big< r^2_{\nu_{\mu}}\, \big> &=& 
2.4\times 10^{- 33}\ \mathrm{cm^2} \nonumber\\
\big< r^2_{\nu_{\tau}}\, \big>&=&
1.5 \times 10^{- 33}\ \mathrm{cm^2}
\label{Numb}
\eea
These values are consistent with various bounds that have appeared 
in the literature
\cite{Salati:1994tf,Allen:1991xn,Mourao:1992ip,Vilain:1995hm,Grifols:1987ed,
Grifols:1989vi,Joshipura:2001ee}.

The theoretical values for 
$\bar{s}_w^{2}(0)$ and $\bar{R}^2(0)$
are obtained from  Eq.(\ref{sweff1}) and 
Eq.(\ref{TransRen})-- Eq.(\ref{RW}).
Since (by construction) these two 
quantities are renormalization-group invariant, one may choose  
any renormalization scheme for computing their value.
In the ``on-shell'' (OS) scheme \cite{Sirlin:1981yz}
the experimental values for the 
input parameters $s_w$ and $\alpha$ are 
$s_w^{\chic{(OS)}}  = 0.231$ 
and $\alpha^{\chic{(OS)}} = 1/128.7$.  
In the same scheme the renormalized 
$\widehat{\Pi}_{\chic A \chic Z}$ is 
given by
\be
{\widehat{\Pi}}_{\chic A \chic Z}(q^2) =
{\widehat{\Pi}}_{\chic A \chic Z}^{0}(q^2) - \Re e\,   
{\widehat{\Pi}}_{\chic A \chic Z}^{0}(M^2_{\chic Z})
\ee
where $\Re e \{...\}$ denotes the real part. Similarly,
\be
\widehat{\Sigma}_{\chic{Z}\chic{Z}}(q^2) = 
\widehat{\Sigma}_{\chic{Z}\chic{Z}}^{0}(q^2) - 
\Re e\,\widehat{\Sigma}_{\chic{Z}\chic{Z}}^{0}(M^2_Z) - 
\, (q^2 - M^2_Z) \Re e\,
\widehat{\Sigma}^{0'}_{\chic{Z}\chic{Z}}(q^2)|_{q^2=M^2_Z},
\ee
where the prime denotes differentiation with respect to $q^2$. 
The closed expressions 
for ${\widehat{\Pi}}_{\chic A \chic Z}^{0}(q^2)$
and $\widehat{\Sigma}_{\chic{Z}\chic{Z}}^{0}(q^2)$ are given 
in \cite{Hagiwara:1994pw}.
Substituting 
standard values for the quark and lepton masses, and choosing for 
the Higgs boson 
a mass $M_H = 150$ GeV, we obtain 
$\bar{R}^2(0) = 1.86\times 10^{-3}/M^4_Z $ 
and $\bar{s}_w^{2}(0) = 0.239$.

\section{Conclusions}\label{sec:Con}

In this paper we have addressed  the observability of the NCR, and its
direct  extraction  from   neutrino  experiments.   The  present  work
constitutes  the  natural continuation  of  the  program initiated  in
\cite{Bernabeu:2000hf},  where  it  was  shown  how  the  NCR  may  be
field-theoretically  {\it defined}  in such  a way  as to  satisfy the
crucial  requirements  of  gauge-invariance and  process-independence.
This  was  accomplished by  resorting  to  the  PT separation  of  the
physical amplitude  involving charged (target)  fermions and neutrinos
into  sub-amplitudes  which  have  the same  kinematic  properties  as
conventional Green's functions, but  are endowed with crucial physical
properties.    The  neutrino   electromagnetic  form-factor   and  the
corresponding   NCR  are  then   defined  through   the  sub-amplitude
corresponding to an (effective) proper vertex.

Our  presentation has mainly  focused on  the following  two important
issues.   First,  in  order  to  assign  an  observable  character  to
individual  sub-amplitudes, in  addition to  the  gauge-invariance and
process-independence    they    must    be   invariant    under    the
renormalization-group.  This point was explained for the first time in
\cite{Bernabeu:2002nw}, and in much  more detail in the present paper.
The requirement  that the  various sub-amplitudes be  individually RGI
resolves a  final theoretical point  related to the definition  of the
NCR, namely  the role and fate of  the universal (flavour-independent)
corrections   stemming  from  the   one-loop  photon-$Z$   mixing.  In
particular,    the   renormalization-group    properties    of   these
contributions  make clear  that  they  cannot form  part  of the  NCR;
instead they are inextricably connected to the flavour-independent RGI
quantity known as the  ``effective'' or ``running'' electroweak mixing
angle.   Second,  we have  shown  that the  NCR  is  indeed a  genuine
observable,  because it  may  be extracted  directly from  experiment.
This has been accomplished by resorting to the neutrino--anti-neutrino
method,   which    allows   the   systematic    elimination   of   the
flavour-dependent box-contributions.   Of course it is  clear that the
processes  considered constitute  thought-experiments, unlikely  to be
realized  in  the  foreseeable  future.  They  do  however  serve  for
clarifying  an  important conceptual  issue,  namely  whether the  NCR
defined through  the PT can be  elevated to the stature  of a physical
observable. This result constitutes the culmination of  
efforts made by various studies presented   
in the literature within the last two decades.  

It is  interesting to notice the absolute  complementarity between the
present work and that  of \cite{Bernabeu:2000hf}.  In particular, once
the NCR has been expressed in terms of physical cross-sections, as for
example  in   Eq.(\ref{sol2})  of   the  present  paper,   its  actual
calculation may  be carried  out in any  gauge-fixing scheme,  with or
without  reference to  the PT,  and it  will always  yield  the unique
answer  of  Eq.(\ref{ncr})  .   On   the  other  hand  ,  without  the
theoretical advancements presented in \cite{Bernabeu:2000hf}, it would
have  been very  difficult to  guess what  the correct  combination of
observables should be.

Now that the observable character  of the NCR has been established, it
would  be interesting  to  undertake  a similar  task  for the  entire
neutrino  electromagnetic form-factor, i.e.   for arbitrary  values of
the momentum transfer $q^2$. One possibility for accomplishing this may
be the detailed study of coherent neutrino-nuclear scattering \cite{JB}. 
We hope to report progress in this direction in the near future.
 
\newpage

\begin{acknowledgments}
This research was supported by CICYT, Spain, under Grant AEN-99/0692.
We thank E.~Masso and A.~Santamaria for various useful discussions, and
D.~Binosi for checking some of the calculations.   
\end{acknowledgments}



\end{document}